\documentclass[journal]{IEEEtran}

\usepackage{amsmath,amsfonts}
\usepackage{algorithmic}
\usepackage{algorithm}
\usepackage{array}
\usepackage[caption=false,font=normalsize,labelfont=sf,textfont=sf]{subfig}
\usepackage{textcomp}
\usepackage{stfloats}
\usepackage{url}
\usepackage{verbatim}
\usepackage{graphicx} 
\usepackage{cite}
\usepackage{breqn} 
\usepackage{amsmath} 
\usepackage{xcolor}
\newcommand{\mv}[1]{{\textcolor{black}{#1}}}
\newcommand{\nv}[1]{{\textcolor{black}{#1}}}
\newcommand{\lb}[1]{{\textcolor{black}{#1}}}

\usepackage{scalerel}
\usepackage{tikz}
\usetikzlibrary{svg.path}

\definecolor{orcidlogocol}{HTML}{A6CE39}
\tikzset{
  orcidlogo/.pic={
    \fill[orcidlogocol] svg{M256,128c0,70.7-57.3,128-128,128C57.3,256,0,198.7,0,128C0,57.3,57.3,0,128,0C198.7,0,256,57.3,256,128z};
    \fill[white] svg{M86.3,186.2H70.9V79.1h15.4v48.4V186.2z}
                 svg{M108.9,79.1h41.6c39.6,0,57,28.3,57,53.6c0,27.5-21.5,53.6-56.8,53.6h-41.8V79.1z M124.3,172.4h24.5c34.9,0,42.9-26.5,42.9-39.7c0-21.5-13.7-39.7-43.7-39.7h-23.7V172.4z}
                 svg{M88.7,56.8c0,5.5-4.5,10.1-10.1,10.1c-5.6,0-10.1-4.6-10.1-10.1c0-5.6,4.5-10.1,10.1-10.1C84.2,46.7,88.7,51.3,88.7,56.8z};
  }
}

\newcommand\orcidicon[1]{\href{https://orcid.org/#1}{\mbox{\scalerel*{
\begin{tikzpicture}[yscale=-1,transform shape]
\pic{orcidlogo};
\end{tikzpicture}
}{|}}}}

\usepackage{hyperref} 

\begin{document}
\title{How to keep pushing ML accelerator performance? Know your rooflines!}

\author{Marian~Verhelst\orcidicon{0000-0003-3495-9263}\,,~\IEEEmembership{Fellow,~IEEE}, Luca~Benini\orcidicon{0000-0001-8068-3806},,~\IEEEmembership{Fellow,~IEEE}, and~Naveen~Verma\orcidicon{0000-0002-8208-5030},,~\IEEEmembership{Senior Member,~IEEE}%
\IEEEcompsocitemizethanks{Received XXX 2024; revised XXX 2024; accepted XXX
2025. This article was approved by Associate Editor XXXXX.}
\IEEEcompsocitemizethanks{M.~Verhelst is with KU Leuven, Belgium and with imec, Belgium. L.~Benini is with ETH Zurich and Università di Bologna. N.~Verma is with Princeton University and EnCharge AI. The work has received funding from the European Research Council (ERC), the Flanders AI Research Program (FAIR), the Swiss State Secretariat for Education, Research, and Innovation (SERI) under the SwissChips initiative and the DARPA OPTIMA agreement no. HR00112490300.}  
\IEEEcompsocitemizethanks{Color versions of one or more figures in this article are available at
XXXX. Digital Object Identifier XXXX }}

\markboth{Preprint of article published in Journal of Solid State Circuits} {Shell \MakeLowercase{\textit{et al.}}: A Sample Article}

\IEEEpubid{0000--0000/00\$00.00~\copyright~2021 IEEE}

\maketitle

\begin{abstract}
The rapidly growing importance of Machine Learning (ML) applications, coupled with their ever-increasing model size and inference energy footprint, has created a strong need for specialized ML hardware architectures. Numerous ML accelerators have been explored and implemented, primarily to increase task-level throughput per unit area and reduce task-level energy consumption. This paper surveys key trends toward these objectives for more efficient ML accelerators and provides a unifying framework to understand how compute and memory technologies/architectures interact to enhance system-level efficiency and performance. To achieve this, the paper introduces an enhanced version of the roofline model and applies it to ML accelerators as an effective tool for understanding where various execution regimes fall within roofline bounds and how to maximize performance and efficiency under the rooline. Key concepts are illustrated with examples from state-of-the-art designs, with a view towards open research opportunities to further advance accelerator performance.
 \end{abstract}


\section{Introduction and Motivation}

\begin{figure}[]
    \centering
\includegraphics[width=\linewidth]{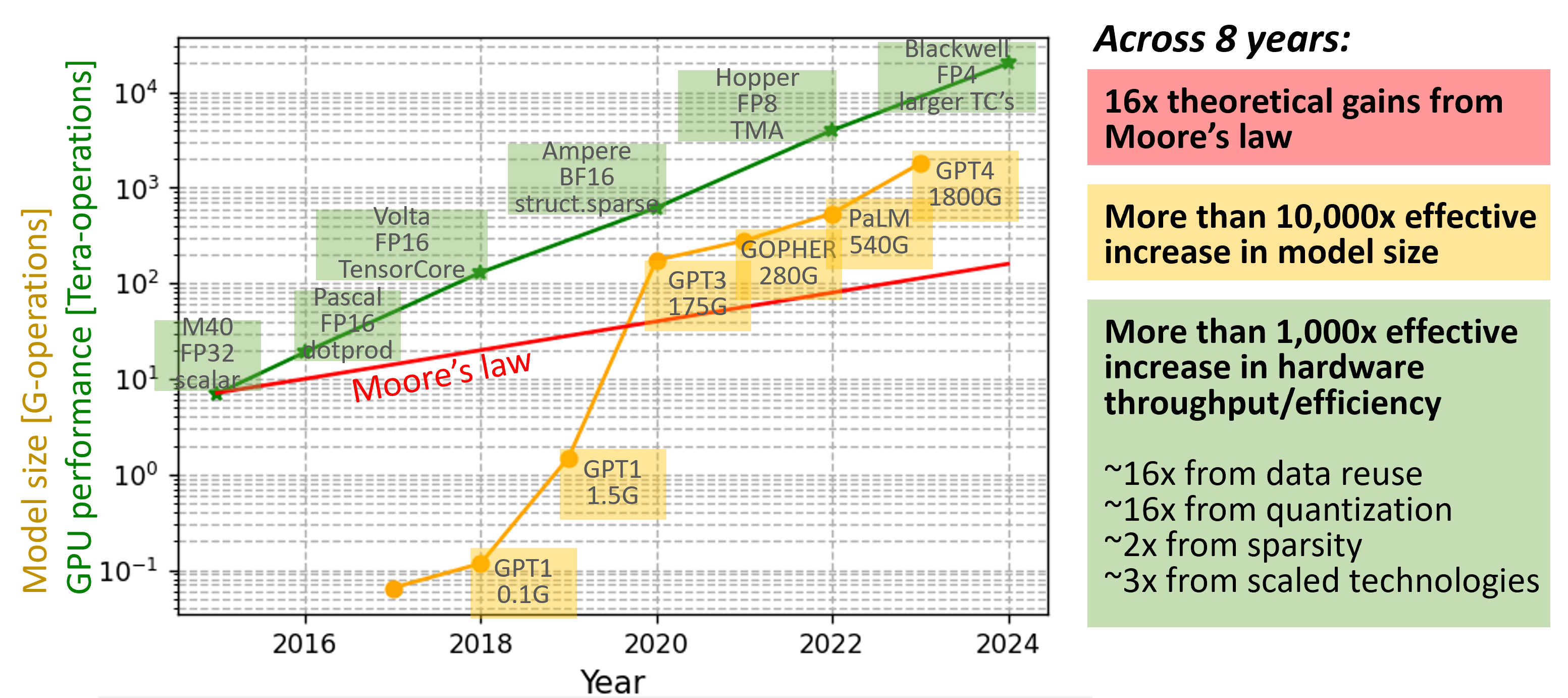}
    \caption{Evolution of ML model size, GPU performance and Moore's law across the last decade. Data from \cite{george2023visualizing} and \cite{dally2024trends}.}
    \label{fig:trends}
\end{figure}

The size of machine learning models has been growing rapidly across the last decade, as illustrated in Figure \ref{fig:trends}. This growth led to an explosion in the required number of operations per ML training or inference tasks, as well as growing memory footprints. The growth rate in complexity of these models substantially outpaces Moore's law (Figure~\ref{fig:trends}). As a result, tracking the requirements in energy efficiency and throughput requires agility in design innovation and clarity around design trade-offs.

This sustained innovation pressure has led to a myriad of dedicated hardware accelerators, exploiting specialized efficiency-boosting techniques, tailored for the ML domain. This push towards specialized architectures is challenged by a growing diversity in workloads. The rapid evolution of ML models, in terms of their topologies (layer dimensions, connectivity), precision requirements, modalities and embedding/de-embedding approaches, etc., require specialization yet without giving up on programmability, necessitating careful balance between flexibility and efficiency.

Industrial and academic research have responded to the challenging needs, achieving impressive nearly 100$\times$ performance improvement every 24 months, maintained over the past 8 years as seen in Figure~\ref{fig:trends}. Despite the wide variety of ML accelerator architectures in \mv{the NPU's (Neural Processing Units) of} commercial products and published in the literature, a large part of their performance improvements can, however, be attributed to different variants of a common list of effective optimizations, including: 
a.) maximizing parallelism; b.) exploiting (spatial and temporal) data reuse; c.) quantization; d.) sparsity; and e.) near- or in-memory compute. 
This paper surveys these techniques, and illustrates their key insights and trade-offs with various examples from state-of-the-art (SOTA) designs.

 \IEEEpubidadjcol

As these techniques are deployed in widely diverging platforms and at different scale, from tiny/extreme-edge devices, to edge processors, to cloud systems, it is difficult to assess them relative to each other. We require a unifying framework to assess their impact and to gain structured understanding of what sets efficiency and performance, as well as what the associated trade-offs are. This paper attempts to do this, leveraging and enhancing the roofline model for throughput and energy efficiency. The goal is to clearly visualize the impact of the different architectural techniques through the roofline representation, thereby enhancing contextual/comparative understanding. Finally, this will lead to insights into fundamental trade-off between the different techniques impacting performance, ultimately leading to open questions and directions for future research.

This paper is organized as follows: Section~\ref{sec:RL} outlines the components contributing to the energy and latency of ML accelerator execution, leading to a visualization comprising two different rooflines. The most impactful efficiency enhancement techniques from recent research are subsequently surveyed and assessed in light of the roofline models in Section~\ref{sec:techn}, illustrated by several examples from recent SOTA. Next, section IV dives deeper into some fundamental trade-offs between the presented techniques, and how they have a conflicting impact on the performance and efficiency rooflines. This will finally lead to an outlook on the future, and directions for further research in Section V.



\section{The roofline models for throughput and energy efficiency} \label{sec:RL}

\subsection{Where does the energy and latency go?}
\begin{figure}[]
    \centering
\includegraphics[width=\linewidth]{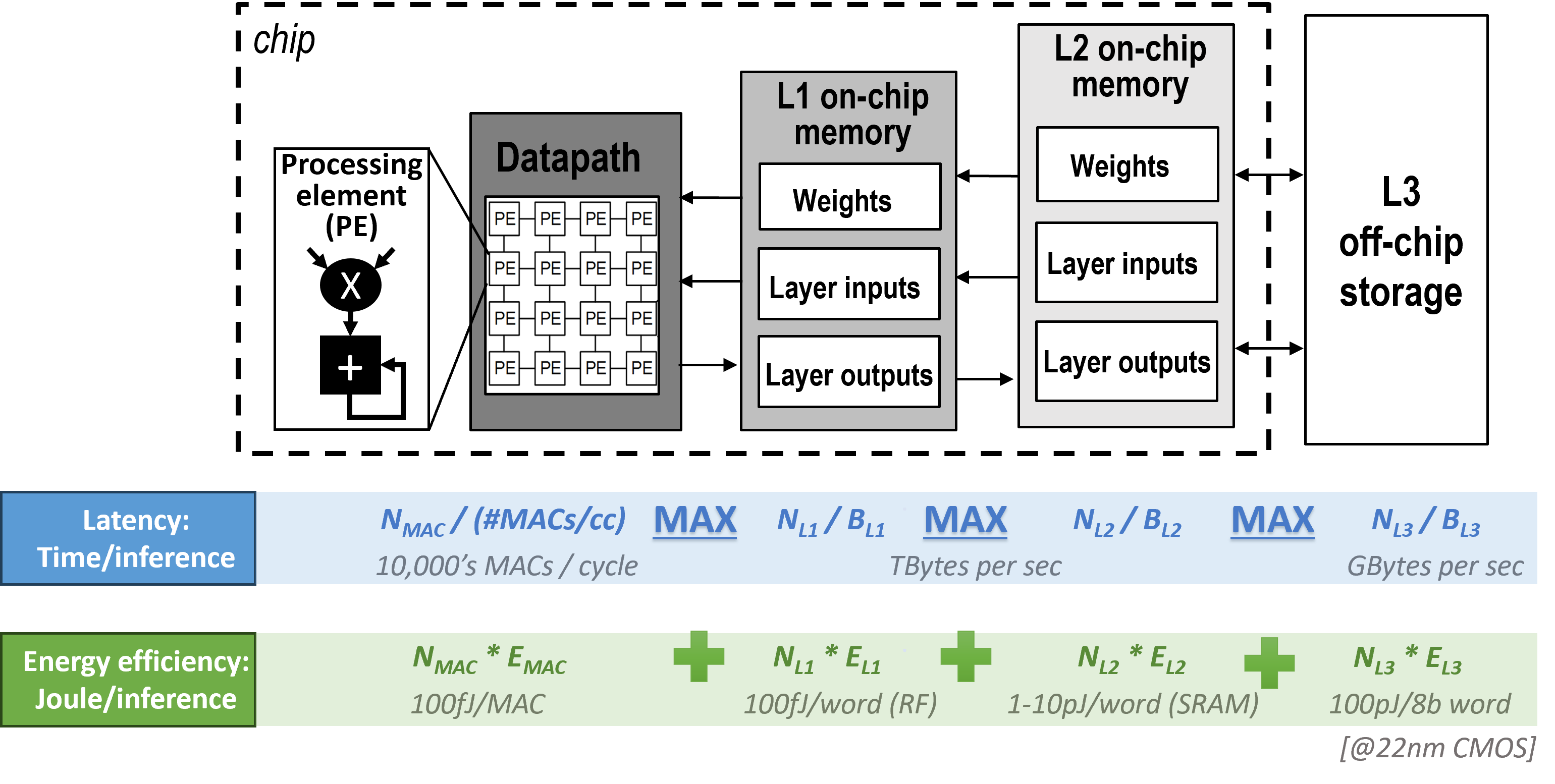}
    \caption{Typical architecture of an ML accelerator. Note that the amount of PE's in the datapath, the number of memory levels in the memory hierarchy and type of memory at each level (registers, SRAM, DRAM) can vary from system to system. Illustrative bandwidth and energy numbers are from a 22 nm technology.}
    \label{fig:energy}
\end{figure}

Almost all ML accelerators, regardless of whether standalone units, data-path extensions to a processor, or tensor cores integrate in a GPU, can be characterized by a common baseline architecture, depicted in Figure \ref{fig:energy}. A regular array of processing elements are each capable of one or more multiply-accumulate (MAC) operations. The data for these operations are fed from an on-chip memory system, typically consisting of a hierarchy of registers and one or more levels of SRAM scratch pads. Additionally, the memory system typically couples to off-chip memory and/or storage subsystems. 

To optimize ML accelerators for throughput and energy efficiency, one has to understand where the energy and latency are spent in such systems. While one could focus on the efficiency of the MAC operation itself, this often only contributes to a portion of the total system energy and latency. Specifically, the energy per inference task $E_{task}$ and latency per inference task $L_{task}$ are heavily impacted by the necessary data movement operations to bring the required data on chip, and to move the data between the different on-chip memory levels.

For a given number of bytes  $N_{Li}$ that must be accessed in each level $Li$ of the memory hierarchy in order to perform a given number of multiply and/or accumulate operations $N_{op}$ per inference, the total energy per inference can be computed as: 
\begin{equation}
E_{task} =  N_{op}.E_{op} + \sum _i N_{Li}.E_{Li}
\end{equation}
with $E_{Li}$ being the energy per byte of memory access at memory level $Li$, and $E_{op}$ being the energy per multiply or accumulate operation.

Latency cannot be derived with a similar sum of the latency impact of the different memory and compute components: As ML accelerators are typically designed to perform compute and data-movement operations in parallel, compute operations can be hidden behind memory transfers or vice versa. As a result, the total latency per inference task is typically obtained by: 
\begin{equation}
L_{task} = \frac{1}{f_{clk}}.max{(\frac{N_{Ln}}{B_{Ln}}, ... , \frac{N_{L1}}{B_{L1}}, \frac{N_{op}}{A_{op}})}
\end{equation}
 with $B_{Li}$ being the bandwidth (bytes/cycle) of memory level $Li$, $A_{op}$ being the amount of arithmetic operators physically on the chip, and $f_{clk}$ being the chip's operating frequency.

To determine which contribution dominates the total energy and latency of an ML accelerator, one can fill in energy and memory bandwidth values of a target technology, as given in Figure \ref{fig:energy} for a 22 nm CMOS. Doing so, it quickly becomes apparent that in both the energy and latency models, the ratios between the operation counts $N_{op}$\footnote{\mv{Throughout the complete paper, only addition and multiply operations are considered towards the number of operations (ops), where a MAC (multiply accumulate) counts for 2 operations.}} and the different memory access counts $N_{Li}$ strongly determine the relative contribution of the different components towards system-level efficiency. In practice, these ratios are determined by the workload and denoted as the "Arithmetic Intensity" $AI$ (also referred to as the "Operational Intensity", or "Compute Intensity") of the target workload for memory level $Li$ \cite{williams2009roofline}: 
\begin{equation}
AI_{Li} = \frac{N_{op}}{N_{Li}} 
\end{equation}
For neural network workloads, the compute intensity typically increases steeply towards the upper levels of the memory stack. A single MAC can have a $AI$ of less than one towards its internal registers, as maximally 1 operand can remain stationary \cite{Eyeriss} and hence multiple words have to be read/written per operation. Yet, higher up in the memory hierarchy, the compute intensity can be drastically higher due to various data reuse opportunities, as will be discussed further in Section III. 

It is important to note that the achievable arithmetic intensities towards the different memory levels are highly network and layer dependent \cite{prashanth2024roofline}. While CONV operators with large input/output tensors might be able to achieve high $AI$'s in the upper level memories, others such as fully connected layers in batch-1 mode, experience low $AI$'s which are rather flat along the memory hierarchy. 
For this reason, it is important to assess performance as a function of $AI$, which will be discussed in the next subsection. 

\subsection{Not one, but two roofline models}
\label{ss:RL}
\begin{figure}[]
    \centering
\includegraphics[width=\linewidth]{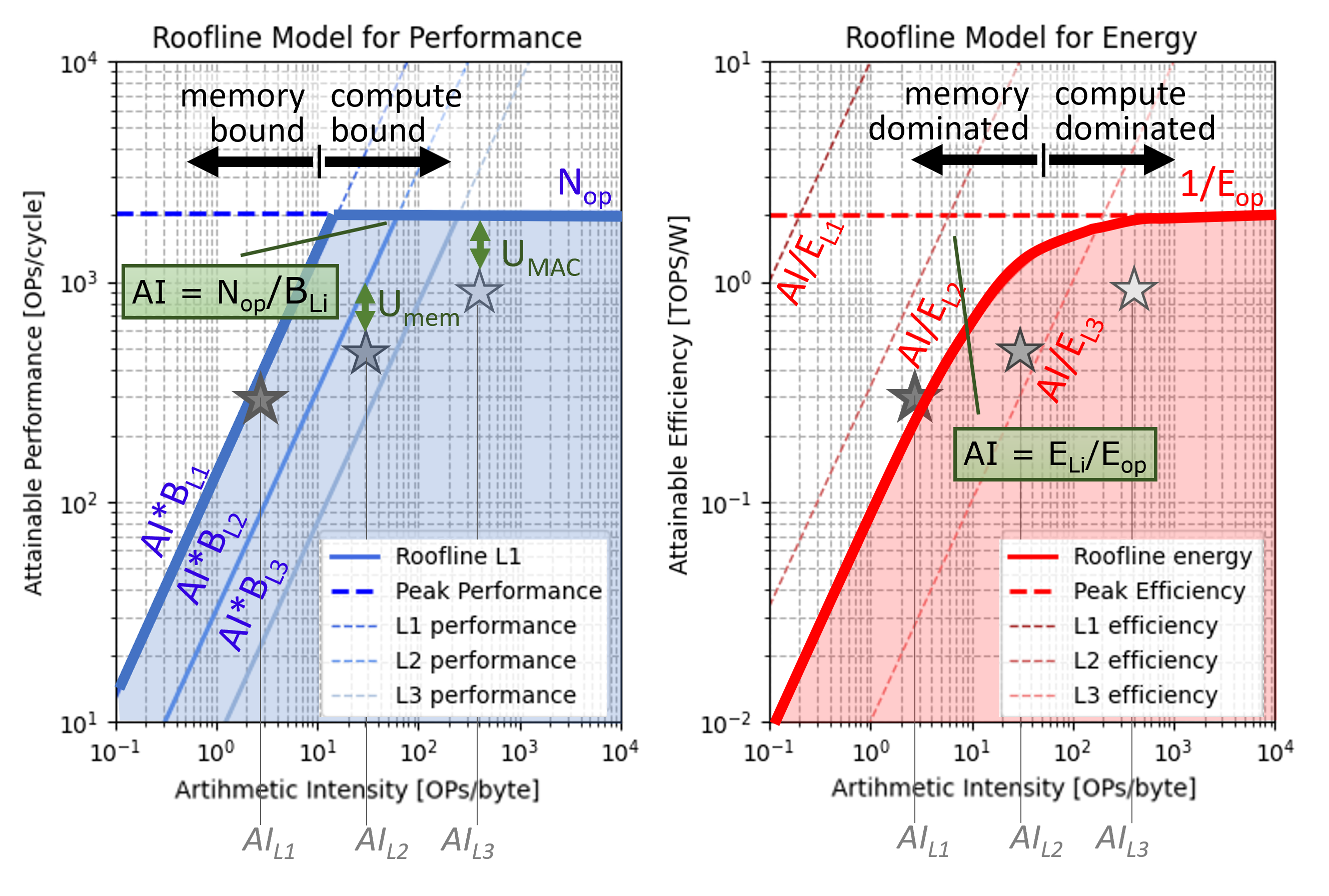}
    \caption{Example roofline models for performance and energy efficiency. \scriptsize{(Assumptions: $N_{op}$=2048, $E_{op}$=0.50pJ/op, $E_{L1}$=0.100pJ/byte, $E_{L2}$=3pJ/byte, $E_{L3}$=100pJ/byte, $B_L1$=128B/cycle, $B_{L2}$=32B/cycle, $B_{L3}$=8B/cycle and for the aggregated energy roofline: $AI$=$AI_{L3}/16$=$AI_{L2}$=$AI_{L1}*16$.)}
    }
    \label{fig:RL}
\end{figure}

To assess throughput across a wide range of potential arithmetic intensities, the computer architecture community introduced the "roofline model" \cite{williams2009roofline}\cite{williams2010roofline}. This model was later also applied to Large Language Model (LLM) inference on GPUs~\cite{yuan2024llm}. It provides a visualization of the attainable processor performance as a function of arithmetic intensity, to quickly assess which system components limit performance for a given workload. The roofline models presented earlier, however, typically focus on the throughput performance of a processor with a single memory level, assuming the $AI$ of the studied workloads remains almost identical across the memory hierarchy, rendering the highest level memory interface the most limiting one. An important distinction in ML accelerators is that the $AI$ of neural networks typically varies widely between memory levels. The opportunities to multi-cast input and weight data, and reduce (aggregate) output data typically cause $N_{L3}<N_{L2}<N_{L1}$ and hence $AI_{L3}>AI_{L2}>AI_{L1}$ (see also Section~\ref{sec:techn}.B and \cite{kwon2019understanding}). 
This requires us to derive the roofline equation for total attainable throughput performance $P_{TP}$ (expressed in operations/sec), starting from Eqn. (2), as:
\begin{equation}
P_{TP} = \frac{N_{op}}{L_{task}} = f_{clk}.min(AI_{Ln}.B_{Ln}, ...,AI_{L1}.B_{L1}, A_{op})
\end{equation}
Visualizing this, results in Figure \ref{fig:RL} (left), in which the memory architecture sets the sloped performance boundary for lower arithmetic intensities, while the MAC compute architecture sets the flat performance boundary for higher arithmetic intensities. Depending on the relative memory bandwidths and arithmetic intensities of the different memory levels, one memory level will be the limiting one and form the overall system roofline (e.g. Level1 in Fig. \ref{fig:RL} (left)). 

While not as common in the literature, it is also valuable to define an \textit{energy} performance roofline \cite{choi2013roofline} for ML accelerators. Energy efficiency is, however, not computed with a min/max operator, but as the sum of the contributing system components (Eqn. (1)), resulting in the following expression for the equivalent Energy performance $P_{E}$, expressed in ops/seconds/Watt or ops/Joule\footnote{\mv{In the remainder of this paper, we will consistently express energy efficiency in terms of TOPS/W $=10^{12} ops/seconds/Watt$, which is equivalent to $ops/pJoule$}}:
\begin{equation}
P_{E} = \frac{N_{op}}{E_{task}} = \frac{1}{E_{op} + \sum_i \frac{E_{Li}}{AI_{Li}}}
\end{equation}
Visualizing this roofline, will no longer give the typical roof shape seen in the well-known throughput curve, yet, will result in a curved roofline, with a low-$AI$ region where memory energy dominates the total efficiency; a high-$AI$ region when compute energy dominates; and a middle region, where the curve is smooth because both components comparably impact total efficiency. The location of the bending point depends on the relative ratio of the different $\frac{E_{Li}}{AI_{Li}}$ terms for the memory levels $Li$.

\mv{It is important to note that the cut off point of both rooflines can lie at very different arithmetic intensities. While the performance roofline has its transition from the memory-bound regime to the compute bound regime at $AI=N_{op}/B_Li$, the energy roofline curves at $AI=E_{Li}/E_{op}$ for a given memory level $L_i$. As both depend on very different parameters, an accelerator can for a certain range of arithmetic intensities simultaneously operate in the compute bound regime for throughput performance, and in the memory-bound regime for energy efficiency. As a result, optimizations can differ significantly depending on whether one cares more for throughput or energy efficiency.}
\mv{Finally note that the throughput roofline only has a sharp transition as long as the compute and memory latency can overlap ($min$-operation in eq. (4)). In case total latency depends on a linear combination of both, also here a rounded roofline curve would appear similar to the energy roofline, as observed in benchmarks in \cite{pellauer2023symphony}.}

While these roofline plots help us in determining the maximum attainable throughput and energy efficiencies for a given $AI$, neither allow us to derive the \textit{effective} efficiency when executing realistic workload. The actual performance will fall below the roofline curve, the extent of which depends on the effective \textit{memory and compute utilizations}, $U_{mem}$ and $U_{MAC}$, respectively. The performance in the flat part of the roofline, will be lowered by the compute utilization $U_{MAC}$, characterizing the ratio of the average number of useful MAC operations performed per cycle, to the total amount of MAC operations available in the ML accelerator. The performance in the sloped part of the roofline, on the other hand, will be lowered by the memory utilization $U_{mem}$, representing the ratio of the average number of useful data bytes accessed from a memory level per cycle, to the peak memory bandwidth at that level.


\subsection{Optimizing performance under the rooflines}

\begin{figure*}[t!]
    \centering
\includegraphics[width=0.95\textwidth]{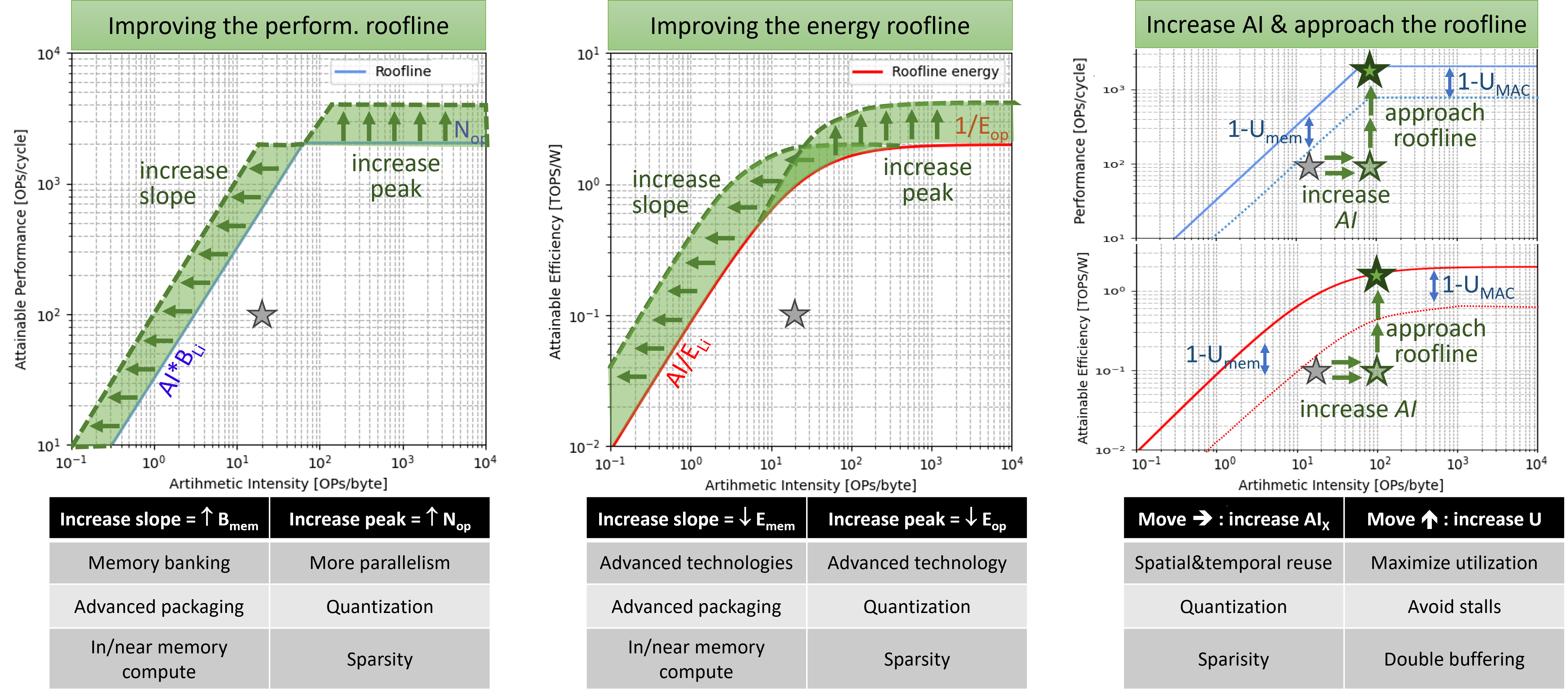}
    \caption{Overview of various architectural and mapping techniques to maximize performance by raising the rooflines and more closely approaching the rooflines.}
    \label{fig:RLimpr}
\end{figure*}

The goal of ML accelerator designers, system architects and workload mappers is to maximize throughput and energy performance. From the discussions in Section \ref{ss:RL} it is clear that this can be done through: (1) raising the rooflines themselves; and/or (2) \mv{by moving to the right, to the compute bound regime and/or (3)} by \mv{more closely approaching the rooflines} (Figure~\ref{fig:RLimpr}). Here, we briefly summarize the most relevant SOTA methods towards these \mv{three} approaches and \mv{visualize their effect in Figure \ref{fig:RLimpr}.} \mv{We will} dive into more detail on each of them in the next Section.

\subsubsection{Raising the rooflines} On the one hand, increasing the compute boundary peak performance (operations per cycle, $N_{op}$) and peak MAC efficiency (\mv{operations per second per Watt}, $1/E_{op}$) allows us to raise the flat part of the roofline. Figure~\ref{fig:RLimpr} outlines various architectural techniques (parallelization for large compute arrays and multi-processors, analog compute, etc.), as well as algorithmic techniques (quantization, sparsity, etc.). On the other hand, increasing the memory boundary in terms of memory bandwidth (bits/cycle, $B_{mem}$) and memory efficiency (energy per access, $1/E_{mem}$) allows us to raise the diagonal part of the roofline and move the knee point more to the left, i.e., to lower $AI$. This, for instance, is possible by optimizing memory architectures for more efficient parallel access, as well as through near/in-memory compute.

\subsubsection{\mv{Move to the compute-bound regime}} 
\mv{As long as workloads operate on the left of the roofline knee point, compute performance and energy efficiency are bounded by the memory bandwidth of the system. To maximally exploit a systems hardware resources, operation should take place as close to the knee point as possible. This typically requires to move right on the horizontal axis,} shifting to larger $AI$ \mv{or in other words doing more compute per fetched data element. One technique to increase AI, is to perform more aggressive quantization, as reducing the precision of the operands increases the number of operations per data byte}. \mv{However, also the hardware architecture can influence the AI. While AI is often considered to only be a function of the considered workload and not of the hardware architecture, we will show in this article that this is not the case for NPU's.} \mv{The physical compute architecture can enable a larger AI by} for example \mv{providing larger on-chip caches or hardware support for} spatial and temporal data reuse, \mv{or by supporting} quantization and/or sparsity \mv{exploitation}. 

\subsubsection{Approaching the roofline} Under a fixed throughput or energy performance roofline \mv{and for a given Arithmetic Intensity}, effective performance can \mv{still be situated well below the roofline. Both in the memory-bounded and in the compute-bounded region effective performance can be limited by utilization. For memory-bounded workloads, this stems from the fact that the memory bandwidth is not used to the fullest. For compute-bounded workloads, this stems from idle cycles or idle MACs on the compute array. In order to maximize performance these utilization losses should be eliminated to move as close as possible to the roofline.} This requires minimizing all efficiency losses stemming from memory and compute under-utilization, \mv{by for example foreseeing support for double buffering to avoid memory stalls or datapath reconfigurability to maximize utilization}. \\ 

\begin{figure}[!b]
    \centering
\includegraphics[width=0.9\linewidth]{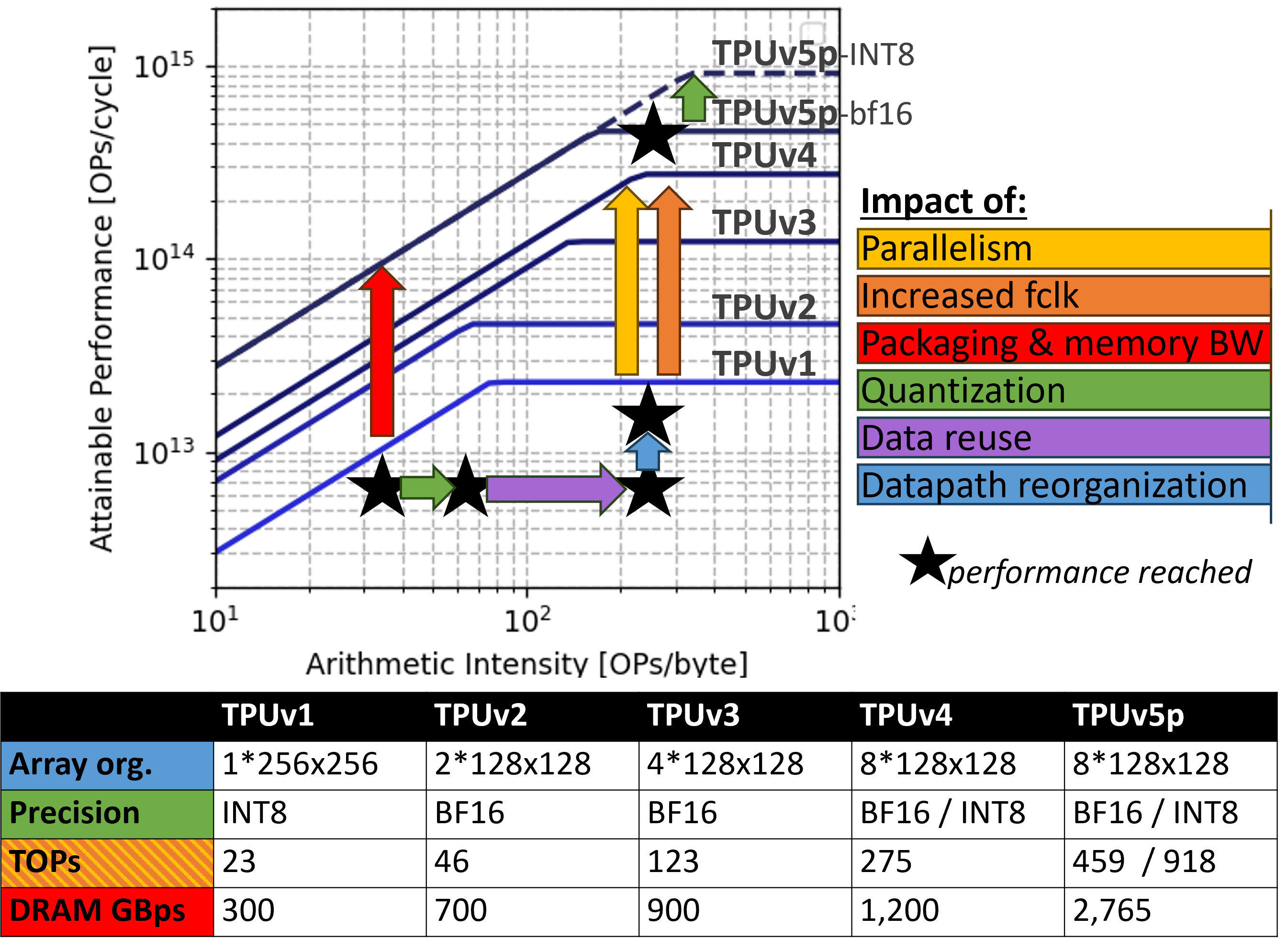}
    \caption{\mv{Evolution of the rooflines of the different TPU generations, together with the impact on Arithmetic Intensity of a typical GeMM workload. Data from \cite{TPU}}}
    \label{fig:TPUroof}
\end{figure}

\mv{Before we dive deeper into those, let's look at how the different techniques come together in a recent NPU evolution. Figure \ref{fig:TPUroof} shows the performance of subsequent generations of Google's TPU. As can be seen, different techniques, such as a wider the DRAM bandwidth, increased parallelism and a higher clock frequency are used to raise the roofline(1). Secondly, hardware support for quantization and data reuse is added to move the workloads to the right of the Arithmetic Intensity axis(2). For example, the TPU makes use of large 128x128 compute arrays for massive spatial data reuse, and is equipped with support for both BF16 and INT8 as of the v4 generation. Finally, hardware is modified to pursue a high utilization in order to operate close to the roofline curve(3). Specifically, the array size is brought from 256x256 to 128x128 to reduce compute utilization losses, while all weight registers are double buffered to reduce memory stalls.}

\mv{In the combination of these different techniques, it is important to always keep an eye on the X-axis location of the throughput and energy knee points. Depending on the AI range of the target workload, one or the other might be more relevant. While improvements of the AI and utilization have impact on the performance under both rooflines, several other architectural enhancements only impact one of the two rooflines. Examples are changes in parallelization ($N_{op}$) shifting the performance roofline, versus changes in operation or memory energy ($E_{op}$ or $E_{Li}$) only impacting the energy roofline.} \mv{After this bird's eye view, let's dive deeper into each of these techniques, to fully understand the opportunities that stem from each of them in light of the performance under the roofline.}

\section{Hardware techniques to maximize performance under the roofline} \label{sec:techn}
This section will dive deeper into various hardware architectural and implementation techniques aiming to increase throughput and/or energy efficiency. 

\subsection{Maximizing parallelism} \label{sec:tech-parallel}
 A direct knob to push the performance roofline up is to increase the amount of parallelism in the ML accelerator. The majority of the neural network operators consist mainly of "multiply-accumulate" (MAC) operations. To raise the flat part of the roofline, the number of parallel multiply-accumulate units in the processing core should hence be increased. Indeed, this trend can clearly be observed over the past decade: From Eyeriss' 168 and Envision's 256 MAC units in 2016 and 2017  \cite{Eyeriss}\cite{moons201714}, over Samsung's and Tesla's $\sim$10,000 of MACs in 2019 and 2020 \cite{park20219}\cite{talpes2020compute}, to today's massively parallel near- and in-memory compute engines with 100,000's of parallel MAC operators \cite{jiao20207, gwennap2020groq, 9365791, 10067422, 10.1145/3543622.3573210}, and often beyond this in the case of multi-core compute engines for providing additional parallelism at the full system level. 
 
\begin{figure}[b!]
    \centering
\includegraphics[width=\linewidth]{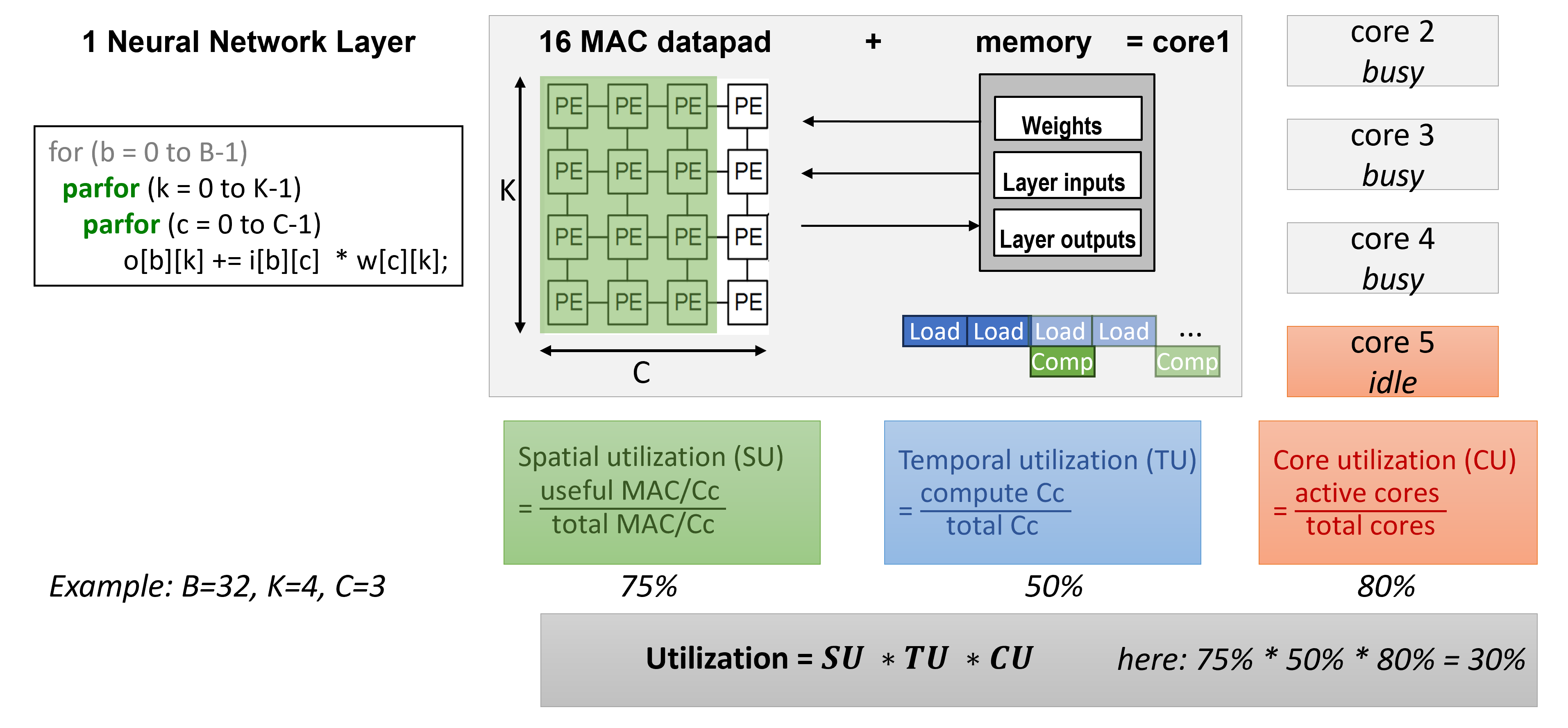}
    \caption{Spatial and temporal under-utilization. Illustration for a 5-core system with an (unrealistically small) MAC array of dimension 4x4 per core, and a MatMul workload of dimensions 32x4x3}
    \label{fig:util}
\end{figure}

Rising to these levels of parallelism, however, comes with important challenges. First, a mapper is required to determine how to spatially and temporally map a target workload across the many parallel processing elements. Assume for example a system with $N$ cores, each consisting of a 2-dimensional array of $M\times M$ MAC operators; and assume as target workload a sequence of $L$ batched neural network layers, each represented as a matrix multiplication (MatMul) of dimension $[B\times C]*[C\times K]=[B\times K]$. This workload can be parallelized across the different cores under various parallelization schemes: \textit{DataParallelism} replicates the network across the cores, and feeds each replication with different data from the batch dimension $B$; \textit{PipelineParallelism} assigns different layers $L$ to different cores, such that they can feed each other in a pipelined fashion; and \textit{TensorParallelism} splits a single layer into multiple chunks along the $C$ or $K$ dimensions, and executes each chunk on a separate processor core. Within a core, the tiled matrix multiplication assigned to a specific core is then further unrolled on to the core's internal $M\times M$ MAC array, for example, spatially unrolling the remainder of the $K$ loop along one dimension and the $C$ loop along another one, while mapping the $B$ loop temporally, as illustrated in Figure~\ref{fig:util}.

It is, however, important to realize that the selected paralellization strategy strongly impacts the efficiency with which the parallel resources are used. This can be quantified in terms of the \textit{utilization} of the processing elements. As detailed in Figure~\ref{fig:util}, the utilization can be broken down into the \textit{core utilization} (ratio of active core count to total core count); the \textit{temporal utilization} (ratio of compute cycles in which the core is not stalled, versus total clock cycles), and \textit{spatial utilization} (ratio of active MAC units versus total MAC units within one processor core). Without proper mapping, the combination of these three utilization factors can quickly lead to severe compute under-utilization, and hence performance far below the theoretical roofline. 

Finally, it is important to realize that the increased parallelism in the cores, raise the roofline of the core, which shifts the knee point of the roofline to higher $AI$'s (see figure~\ref{fig:RLimpr} (left)), making the system more likely to be memory bound, potentially leading to memory-induced reduction of the temporal utilization ($TU$
). Indeed, the execution of a large number of MAC's per clock cycle, also demands a large data volume to be dragged into the compute array per clock cycle. Although theoretically this memory pressure can be resolved by increasing memory bandwidths across all levels of the memory hierarchy to shift the diagonal part of the roofline up, this is often not feasible in practice. ML accelerators often already maximally exploit distributed register files, widely-banked SRAM memories, and high-speed (costly) off-chip DRAM, whereby further bandwidth increases come at substantial area and energy costs. 
Fortunately, it may be possible to increase parallelism without having to proportionally increase memory bandwidth, by employing mappings that leverage increased data reuse, when/if the workloads allow this.

\subsection{Exploiting spatial and temporal data reuse}
The operands of the MAC operations in neural networks have a potential data reuse factor which can easily go up to 10,000's. For example, the weights in a Conv layer or the operands of a large MatMul operation can be fetched once, and subsequently reused in many dotproduct operations. With some exceptions (e.g. weights in a fully connected layer with batch size 1, or inputs in a depthwise layer), neural network compute kernels thus exhibit large intrinsic arithmetic intensity. This, however, does not automatically result in operating at the top-right corner of the roofline model. The processor hardware has to be able to exploit the kernels' data reuse opportunities at the datapath level. 
This involves the alignment of the execution schedule to the hardware implementation, in particular to flexibly leverage both spatial and temporal data reuse as will be explained next. \\

\textit{Spatial data reuse} refers to reusing data within a given clock cycle across more than one MAC unit in the datapath. 
\begin{figure}[]
    \centering
\includegraphics[width=\linewidth]{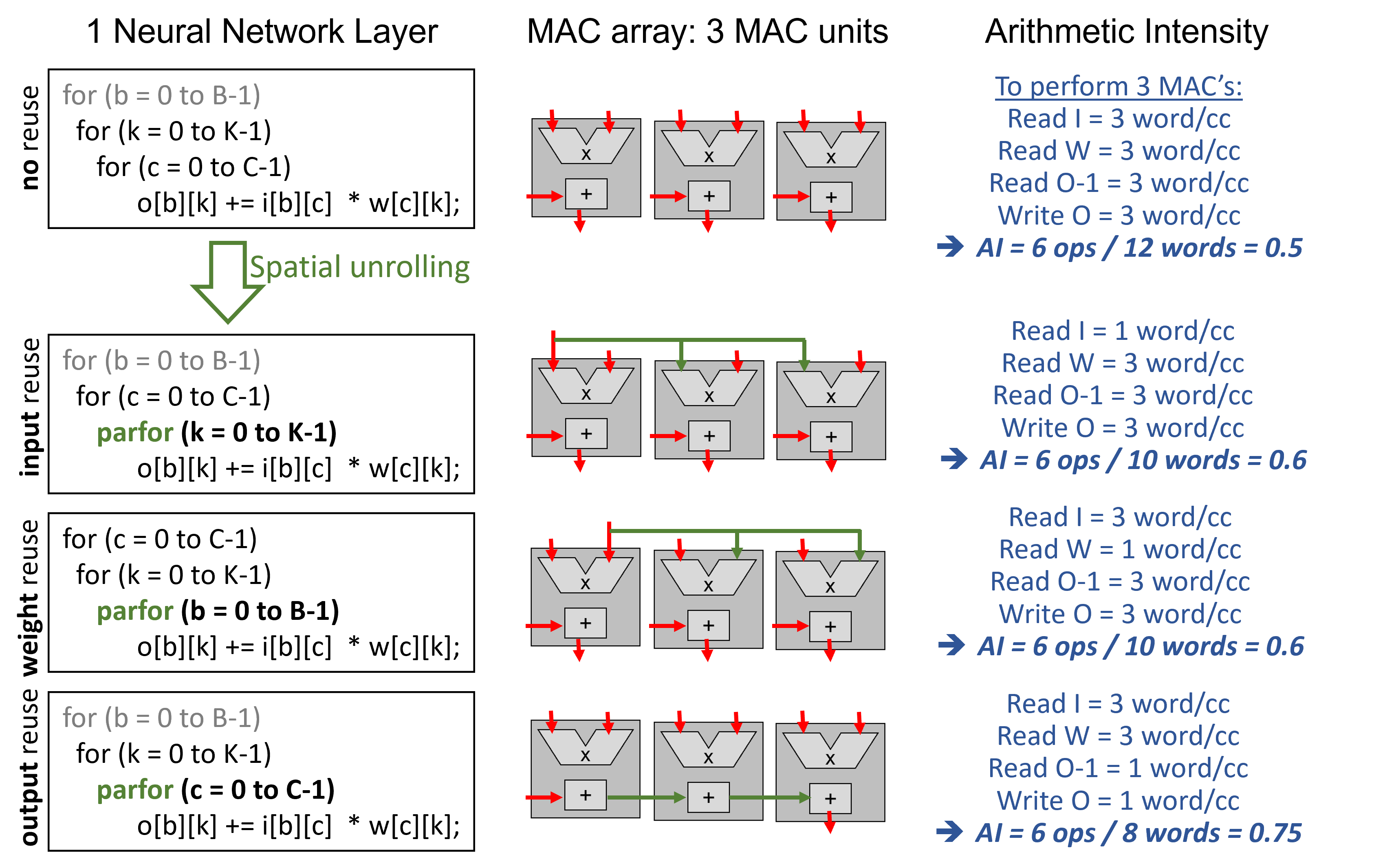}
    \caption{Spatial data reuse concepts}
    \label{fig:spreuse}
\end{figure}
Figure~\ref{fig:spreuse} illustrates how parallel MACs in a MAC array can spatially exploit weight, input, or output data reuse, to increase the effective arithmetic intensity. This requires hardware modifications, to enable direct data sharing paths between the hardware units. Without such provisions, the inherent arithmetic intensity of the compute kernel cannot be spatially exploited at the datapath level. All SOTA ML processors exploit such spatial data reuse along one or more operand dimensions of their datapath, typically consisting of hundreds to thousands of MAC units. The examples of Figure~\ref{fig:spreuse} all only contain a single \textbf{parfor}-loop (=spatially parallel for-loop), corresponding with one-dimensional spatial data reuse. This can be found in vector processors, which are typically equipped with a Fuse Multiply Add (FMA) instruction, exploiting output reuse. 

\begin{figure}[]
    \centering
\includegraphics[width=0.8\linewidth]{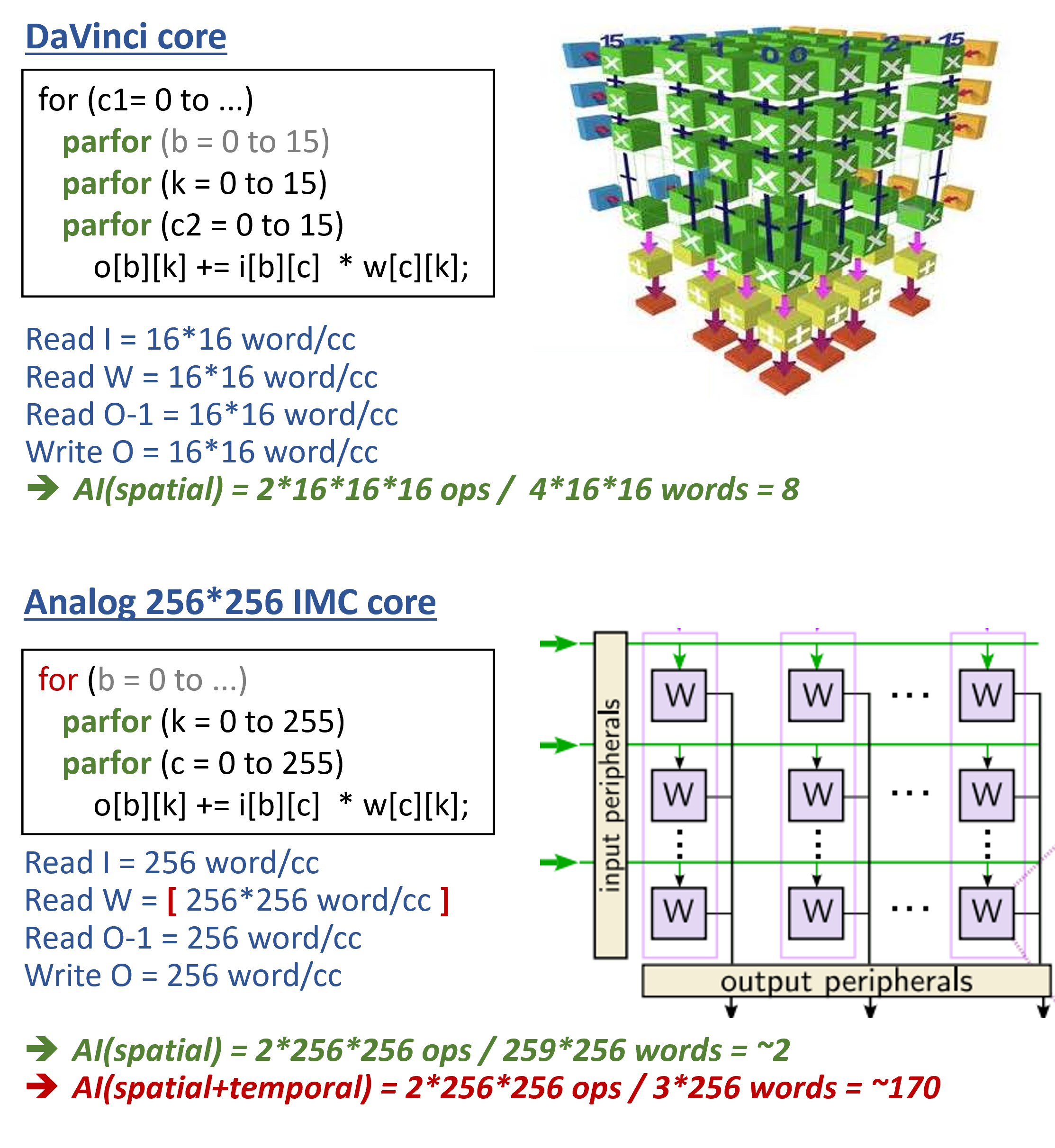}
    \caption{Spatial data reuse within the DaVinci core\cite{liao2019davinci}, and an analog IMC core with 256 rows and columns.}
    \label{fig:davinci}
\end{figure}

However, $AI$ typically does not improve strongly from reuse across a single operator dimension, as the memory accesses of the other operands will quickly start to limit reuse. As a result, ML processors are commonly equipped with 2- or even 3-dimensional datapaths, and hence multiple spatial parfor-loops in their dataflow representation, which jointly exploit different forms of the three spatial reuse concepts along each MAC array dimension. The Envision processor \cite{moons201714}, for example, combines 16-way input reuse and 16-way weight reuse, while the Huawei DaVinci cores \cite{liao2019davinci} or Nvidia Tensor Cores \cite{markidis2018nvidia}
exploit all three forms in a 3-dimensional MAC array, as illustrated in Figure~\ref{fig:davinci} (top). 

Such multi-dimensional spatial reuse, reduces the memory pressure from being linearly dependent on the number of MAC elements $A_{op}$, to only scaling with $\sqrt[3]{A_{op}}$, substantially increasing arithmetic intensity. More importantly, the larger the compute array is, the more spatial data reuse can be accommodated. As a result, larger compute arrays can typically operate at higher arithmetic intensities in the compute-dominated region of the energy roofline (Figure~\ref{fig:RL} (right)), which represents the most energy-efficient region of operation. 
Such very large and efficient compute arrays can be found in recently emerging in-memory compute architectures. Typically these architectures (Figure~\ref{fig:davinci} (bottom)) only exploit spatial data reuse across 2 dimensions: While they can massively reuse inputs and output data along the memory rows and columns, each weight is used only once each clock cycle. As long as the weights do not have to be reloaded, this allows them to operate at very high arithmetic intensity. Loading new weights every clock cycle, however, would drastically limit the spatial $AI$ benefits, as numerically illustrated in Figure~\ref{fig:davinci} (bottom). This is, however, avoided through temporal data reuse, discussed next. \\ 

\textit{Temporal data reuse} can further increase the effective arithmetic intensity. Specifically, it reduces the number of data accesses per compute operation, not by reusing data within one clock cycle (spatial data reuse), but by reusing data \textit{across} clock cycles without repeated memory fetching. When studying the nested for-loops embedding the ML MAC operations (e.g. see Figure~\ref{fig:spreuse} and \ref{fig:davinci}), it is clear that for each for-loop one operand remains constant across different iterations of the for-loop. For example, in the analog IMC core (Figure~ \ref{fig:davinci} (bottom)) the nested for-loop along the $"b"$ dimension reuses the same weight element $w[c][k]$ across all its temporal iterations, as the operand is not dependent on the index $[b]$. This allows us to keep the weights \textit{stationary} across the for-loop iterations, also denoted as a \textit{"weight-stationary" dataflow}. Such weight-stationarity is exploited in almost all in-memory compute cores \cite{wang2023charge, papistas202122, valavi201964}. 

Beyond weight-stationary, processors can also be input- or output-stationary, together forming the three orthogonal forms of temporal data reuse. When describing processor dataflow in terms of their nested "for"- and "parfor"-loops, one can easily derive the stationarity of the processor under study. The lowest for-loop above the spatial "parfor" loops determines the temporal data reuse towards the registers. For example, in the DaVinci-example (Figure~\ref{fig:davinci} (top)) the lowest for-loop is a $"c"$-loop, and hence irrelevant for the output operand. This represented dataflow is hence an \textit{output-stationary dataflow}, which is very common in fully digital ML processors. Also the aforementioned Envision chip \cite{moons201714}, Tensor cores \cite{markidis2018nvidia} or Tesla NPU \cite{TeslaNPU} are output stationary processors. As the outputs typically require a larger word length, and hence memory bandwidth, compared to weights or inputs, output stationarity(=temporal output reuse) has a significant impact on a processor's $AI$.

\begin{figure}[]
    \centering
\includegraphics[width=0.7\linewidth]{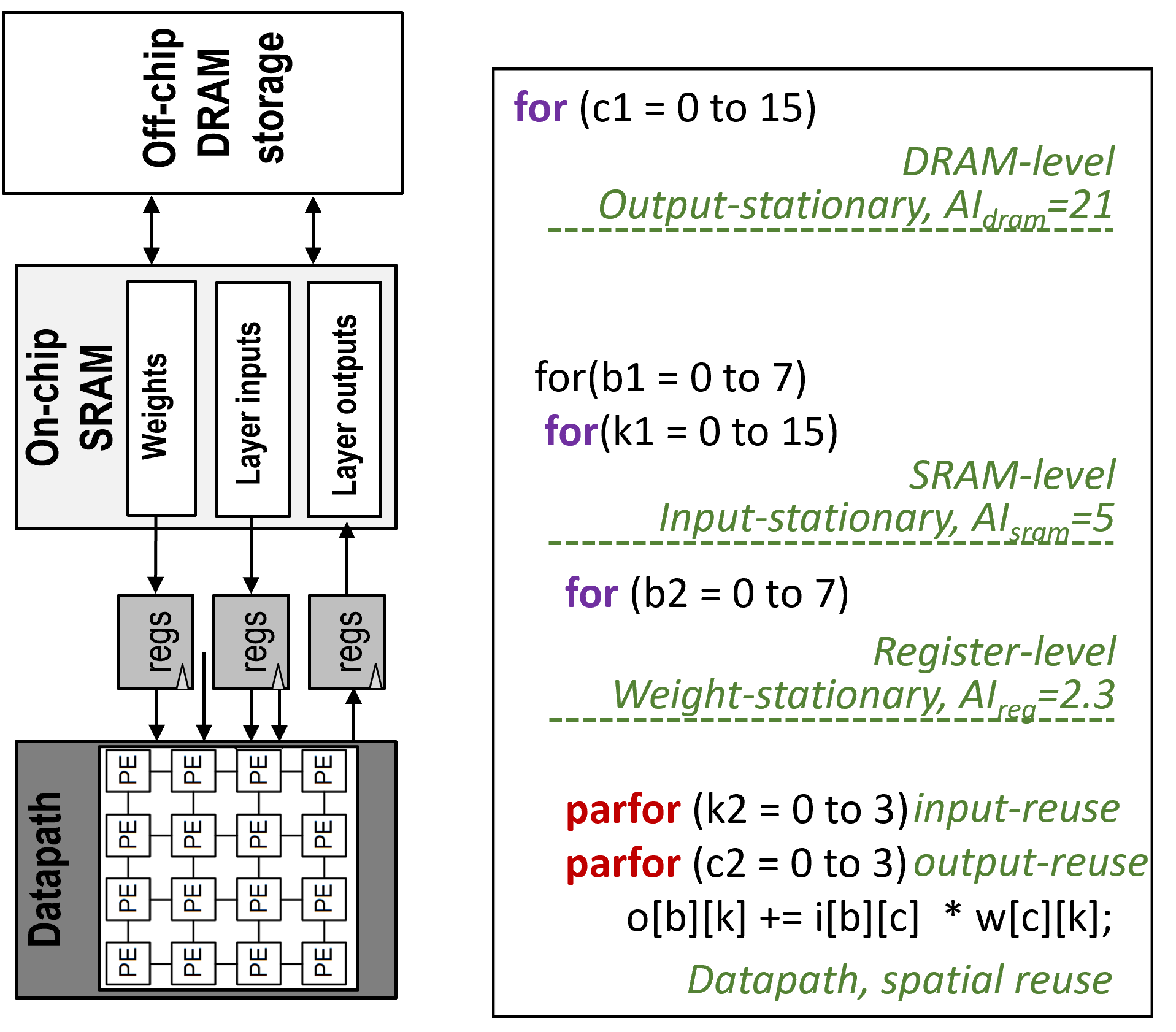}
    \caption{Nested for-loop representation of workload mapping and tensor allocations across the memory hierarchy.}
    \label{fig:allocation}
\end{figure}

In contrast to spatial data reuse when multiple reuse dimensions can be combined, only a single stationarity dimension can be exploited per memory level. Indeed, 
it is not possible to keep two operands constant across subsequent MAC computations, as this would lead to repeated, redundant operations. It is, however, possible, to exploit different forms of temporal data reuse across the memory hierarchy. Executing larger tensor operations in a core typically involves tiling the input, weight, and output tensors, corresponding to splitting/blocking the different for-loops, and allocating them across different memory levels, as done in Figure~\ref{fig:allocation}.

Here it is clear that each level of the memory hierarchy can exploit a distinct form of stationarity, as the lowest for-loop tiled within a memory level determines which tensor can remain stationary towards the lower level memory. This also explains why the arithmetic intensity increases when moving from the L1 to L2 and L3 memory level.  Optimal combinations of such spatial and temporal unrolling are therefore indispensable towards packing as much as compute for a given memory bandwidth.

\subsection{Exploiting quantization}
\label{sec:quant}
Quantization is currently the most actively exploited technique to boost efficiency in ML accelerators. A quantitative estimate of the major performance improvements achieved by GPUs thanks to quantization (16x), has been reported by Nvidia \cite{dallyHC2023} and is confirmed by extensive literature surveys covering a wide range of accelerators \cite{reutherHPEC2023,NPUComparison,yuan2024}.

From a hardware-centric viewpoint, quantization is easily exploitable and its effect on the roofline, \lb{as shown in Figures \ref{fig:roofline-opt} and \ref{fig:TPUroof}}, is to raise the achievable peak compute -the height of the compute-bound region - as more operations can be performed faster in the same silicon area, and to increase the arithmetic intensity of a ML workload, as more operations are performed for each byte exchanged with memory. Energy improvements are potentially superlinear, due to the fact that key operators in ML workloads, such as MAC units, have time and space complexity that is super-linear on the bit-width on their input operands \cite{bertaccini2024}. 

Hence, it is not surprising that the hardware design community has been aggressively pursuing quantization. Recent research results have demonstrated that digital compute efficiency as high as \mv{POPS/W} can be achieved with extreme, ternary and binary quantization \cite{scherer2022,moons2018}.  On the other hand, such an extreme quantization comes with significant model accuracy loss. The current production models can be compressed down to 4-bit coefficients and 8-bit activations, but algorithmic research is actively working to lower these thresholds \cite{wangbitnet2023}.

\begin{figure}[]
    \centering
\includegraphics[width=\linewidth]{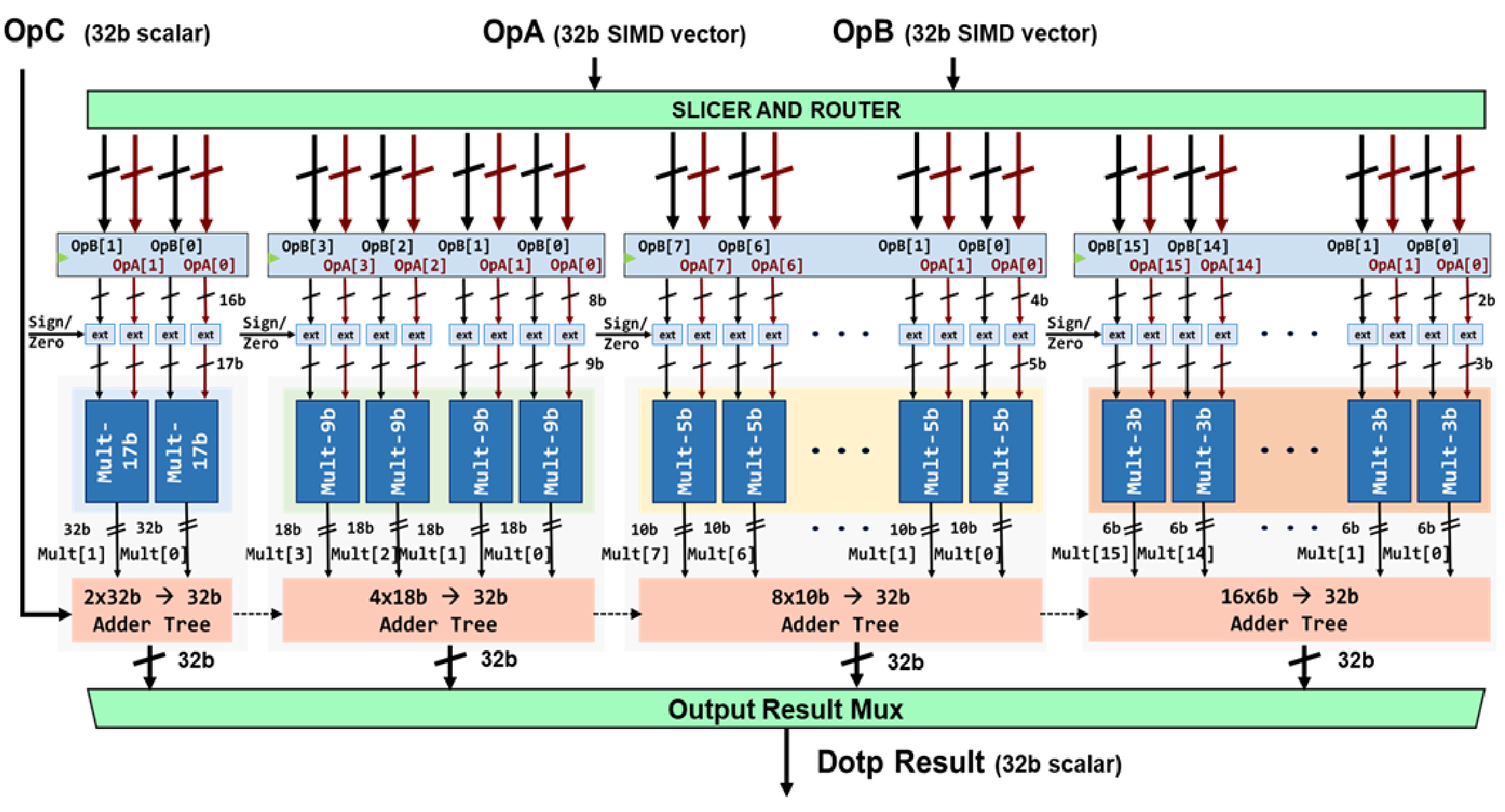}
    \caption{RISC-V mixed-precision SIMD dot-product unit}
    \label{fig:simd}
\end{figure}

Hardware support for quantization in instruction processors, through dedicated instruction set architecture (ISA) extensions, has been extensively explored, leveraging the well-known single-instruction multiple-data (SIMD) approach. As an example, the RISC-V ISA has been extended to support all mixed-precision combinations of power-of-two subsets of 32-bit registers (namely 2, 4, 8, 16 bit operands) \cite{nadalini2023,contimarsellus2024}, as well as binary and ternary operands.  Figure \ref{fig:simd} shows the design of a SIMD  {\it expanding}\footnote{this denomination stems from the fact that the accumulator is computed and stored at higher precision that the operands, 32bit in this case.} dot-product unit for a 32-bit RISC-V core, supporting all power-of-two mixed precision quantized operands below 32-bit. The slicer\&router block splits data coming from the $32b$ input registers in the appropriate narrow operands and routes them the MAC units in the dedicated data-path section. Sign-extension is then performed for the smaller operand; Multiplication is performed in parallel on 8 $5b\rightarrow10b$ multipliers, followed by the $8\times10b \rightarrow 32b$  adder tree. Results are stored in the $32b$ accumulator register.  Note that distinct dot-products units are instantiated for each power-of-two (plus sign extension) bit-width to allow operand isolation and reduce switching activity. 

\begin{figure}[]
    \centering
\includegraphics[width=\linewidth]{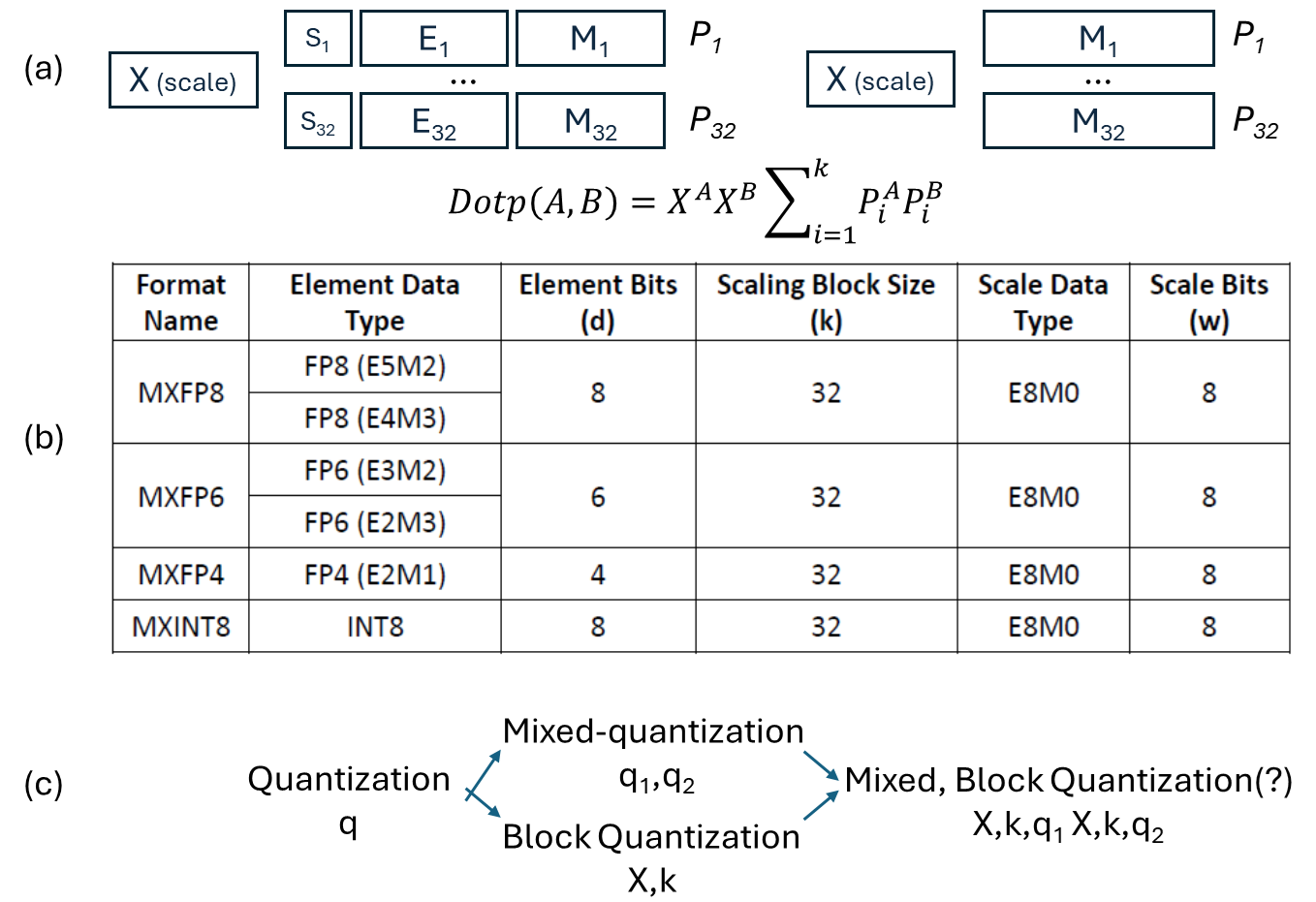}
    \caption{(a) Block quantization and dot-product of block-quantized vector fragments, (b) OCP microscaling specification of block-quantized numbers, (c) from basic quantization to mixed and block quantization, and future block and mixed quantization}
    \label{fig:quant}
\end{figure}

The SOTA hardware support for quantization is moving along two main axes of evolution: First, fine-grain control of mixed quantization, where large multioperand operators, such as large multiplier arrays followed by adder trees for dot-product computation, are designed to enable different and configurable bitwidth for different classes of operands (activations vs. weights) \cite{dallyHC2023}. Second, {\it block-quantization} support is getting increasingly emphasized, as operand-by-operand (scalar) quantization has plateaued \cite{rouhani2023}.  Block-quantized data formats for ML are also being standardized in an effort to facilitate model portability across different hardware platforms: the Microscaling standard formats \cite{rouhani2023}, proposed by Nvidia, AMD, Intel, Microsoft, Meta, Arm, and Qualcomm in the Open Compute Project (OCP), are specified down to 4-bit floating-point for blocks 32 or 64 numbers, with a shared 8-bit exponent.  Figure \ref{fig:quant} provides a visual summary of block quantization as standardized by OCP and depicts the trajectory from simple scalar and uniform quantization to advanced mixed and block quantization schemes. \lb{The most recent GPU generation from NVIDIA, Blackwell, supports all these formats, including the aggressively scaled MXFP6 and MXFP4 \cite{tirumala2024}}.

\begin{figure}[]
    \centering
\includegraphics[width=\linewidth]{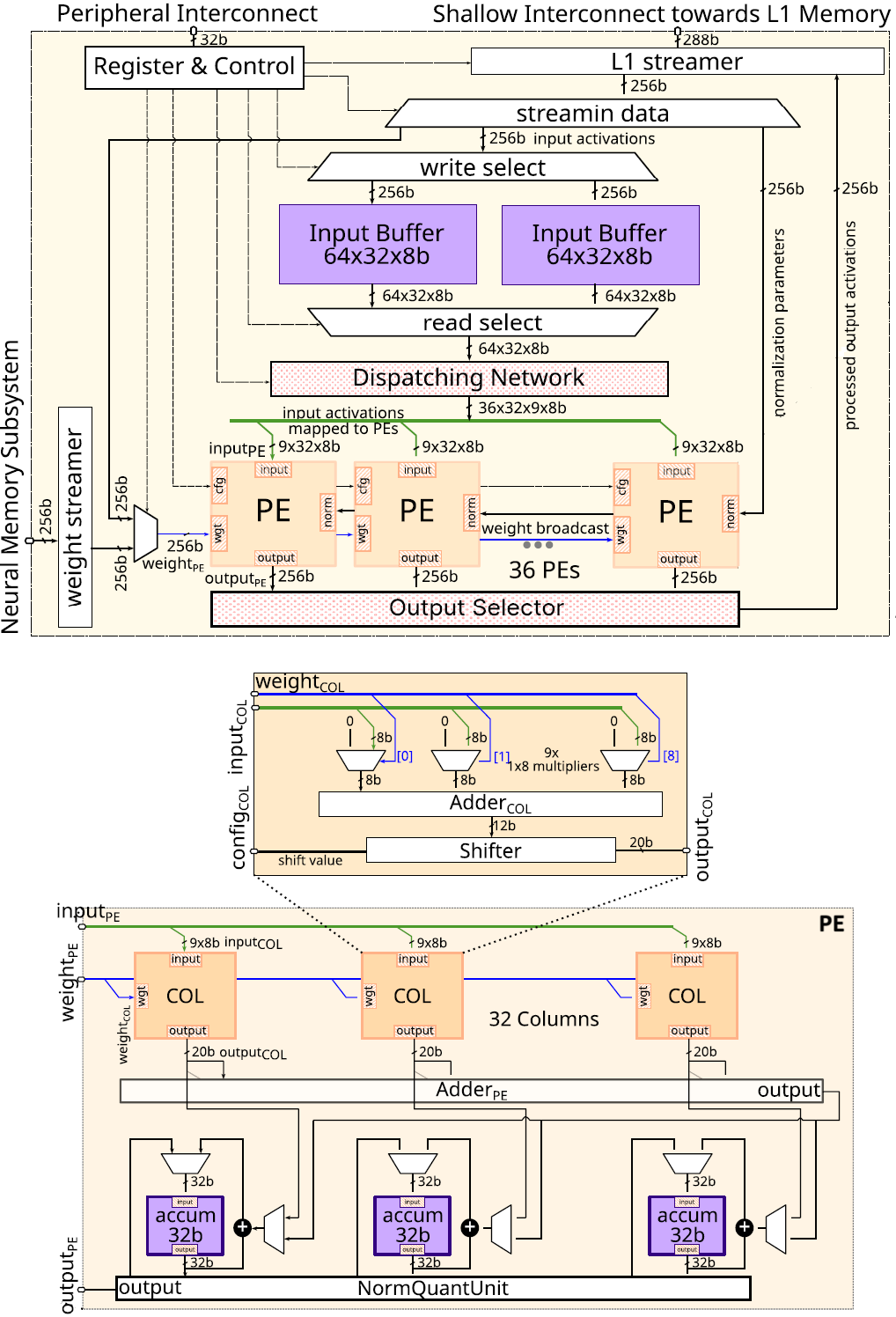}
    \caption{Neureka mixed-precision accelerator architecture design}
    \label{fig:neureka}
\end{figure}

A representative design of a specialized tensor accelerator with finely-tunable block quantization support for weights is Neureka \cite{prasad2024} shown in Figure~\ref{fig:neureka}. Activations and weights are read in blocks through wide ports from L1 memories (not shown in figure). Blocks are buffered in internal (L0) latch-based register files, which are preferred over SRAMs because required capacity is low and energy-per-access can be minimized more aggressively than with SRAMs by voltage scaling, which can be pushed further for logic cells than for SRAM macros.  The datapath is designed to access bit slices of each weight blocks. Multiplication of bits of the weights, with 8-bit activations is performed, one bit at a time, and products are accumulated into multi-bit numbers (12 bits), which are then locally shifted over a 20-bit range. The shifted output of each COL combinational datapath is then further accumulated by a larger adder tree that collects multiple local shifted sums. Results are then stored into output accumulation registers in an output-stationary dataflow.

\begin{figure}[]
    \centering
\includegraphics[width=\linewidth]{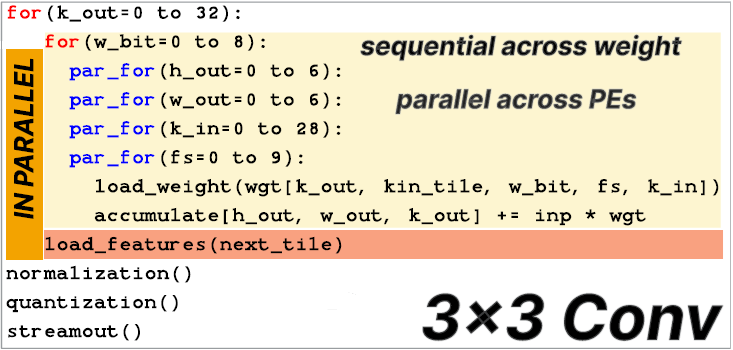}
    \caption{Nested loop for execution of a $3\times3$ convolutional layer in Neureka. Note that shift-and-add loop is executed sequentially bit-by-bit for the 8bit weights}
    \label{fig:loops}
\end{figure}

This datapath can handle  weights quantized between 2 and 8 bits. The number of  cycles  needed to compute the final output-stationary summation grows linearly with $N_w$, where $N_w$ is number of bits used to represent weights. Weight precision is temporally unrolled, and activation precision is spatially unrolled. The bit-by-bit shift and accumulate steps can be seen as one additional dimension in the nested for-loop notation discussed in section \ref{sec:techn}: this is shown in Figure \ref{fig:loops}. In this example, Neureka's datapath accumulates in parallel over inputs and output channels and convolutional window size, but sequentially over the bit of the weights. 

Note that Neureka does not offer tunable activation quantization: this extra degree of freedom has also been explored in the literature \cite{contimarsellus2024}, but quantization of activation below 8 bits is usually not accuracy-neutral. Closing this accuracy gap at a high level of robustness is an active research field. 

An implementation of Neureka in 16nm TSMC technology, featuring 36x32 COL units achieves end-to-end (including L1 memory access cost) energy efficiency ranging from 8.84 \mv{TOPS/W} (2-bit weights) to 2.68 \mv{TOPS/W} (8-bit weights) at 0.6V \cite{prasad2024}. Note that energy scaling is not perfectly linear in the number of bits, because a fraction the execution cycles does not scale with weight bitwidth: a few cycles are needed for scale and offset related computation. Nevertheless, significant energy efficiency boost is achieved ($3.5\times$ when scaling  weight bitwidth by a factor of 4), by aggressively quantizing weights.

\subsection{Exploiting Sparsity} \label{sec:sparsity}

Sparsity is present in most classical and advanced ML models, as coefficients tends to cluster around zero after training, As a consequence, activations are also pulled close to zero, albeit with a smoother distribution. Sparsity exploitation is also one of the key efficiency boosters for incumbent GPU architectures \cite{dallyHC2023}, although to a lesser degree compared to quantization (2x vs. 16x). On the other hand, the impact of sparsity is likely to grow as the trend is toward large-scale models that are intrinsically highly sparse, not just as a result of training. {\it Mixture-of-experts} models\cite{yuan2024} are notable examples of this trend. 

\begin{figure}[]
    \centering
\includegraphics[width=\linewidth]{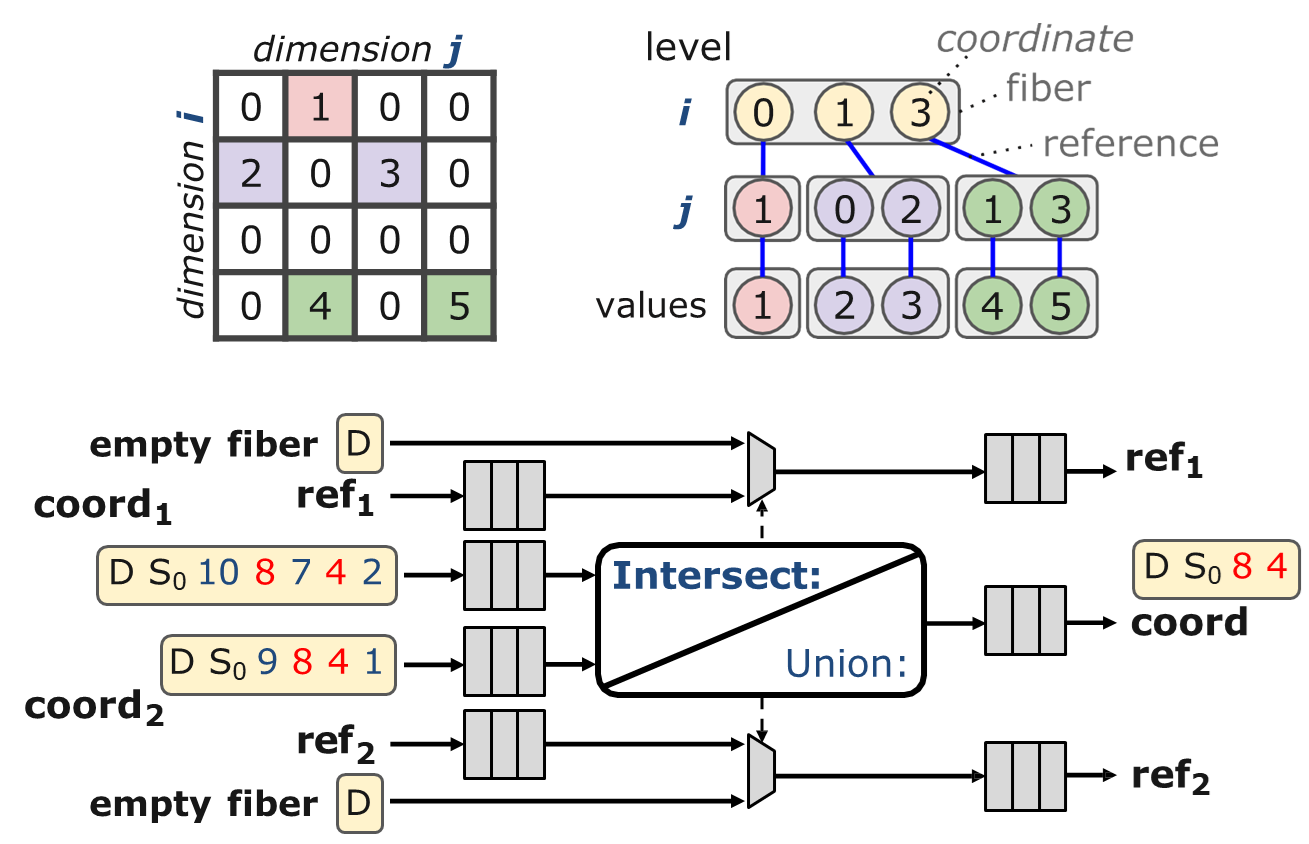}
    \caption{Onyx's fibertree sparse tensor format and hardware unit for intersection and union operators \cite{koul2024}}
    \label{fig:onyx}
\end{figure}

Hardware support for extremely sparse models with non-structured sparsity is being explored, as the rapidly growing model size trend may soon impose the use of indexed data structures to handle high degrees sparsity for the largest "leadership" models.  An example of accelerator with hardware support for highly sparse tensor computation is the Onyx coarse-grained configurable array (CGRA) architecture \cite{koul2024}. The sparse data structure supported by Onyx is known as \emph{fibertree}, and it is depicted at the top of Figure \ref{fig:onyx}. The fibertree is a two-dimensionally-indexed data-structure which indexes non-empty rows (or columns) and for each one of them it indexes the respective non-zero elements. Only indexes and non-zero data elements are stored in memory, thereby achieving large compression factors if sparsity is high and non-structured.

For efficient execution in the Onyx accelerator, the fibertree is encoded, as shown in the bottom of Figure \ref{fig:onyx}, in streams of indexes, read sequentially by the hardware engine, which ingests parallel streams for the rows/columns of the tensors involved in the operation being targeted. In the case of the intersection datapath, used for instance for computing the dot products of rows with columns needed for matrix multiplication, only equal indexes in both streams ($coord_1$ and $coord_2$) are output by the intersector ($coord$ stream), and the corresponding data elements are then loaded from memory for product calculations, which are then locally accumulated. 

While the reduction of memory bandwidth to access data is apparent (only the strictly needed data elements and their indices are loaded), access to the non-zero elements has lower block locality compared to dense tensors; hence efficiently loading data from remote, off-chip DRAM memories, which generally benefit from burst access to blocks of continuous data, is a challenge. Thus, most hardware optimization efforts have focused on {\it structured sparsity}. $N:M$ sparsity is widely supported, where a maximum of $N$ nonzero elements is enforced over an array of $M$ elements \cite{zhou2021learning}. $N:M$ sparsity is hardware friendly, especially if $N$ and $M$ are design-time constants: indexes and data have constant size and efficient block transfers from/to memory remain viable. As a result many production ML acceleration engines, most notably GPUs, support $N:M$ sparsity \cite{choquette20213} (only 2:4 sparsity is supported natively in hardware). On-going efforts are exploring more advanced structured sparsity models, for instance, multidimensional sparsity for tensors \cite{castro2023venom}. 

\begin{figure}[]
    \centering
\includegraphics[width=\linewidth]{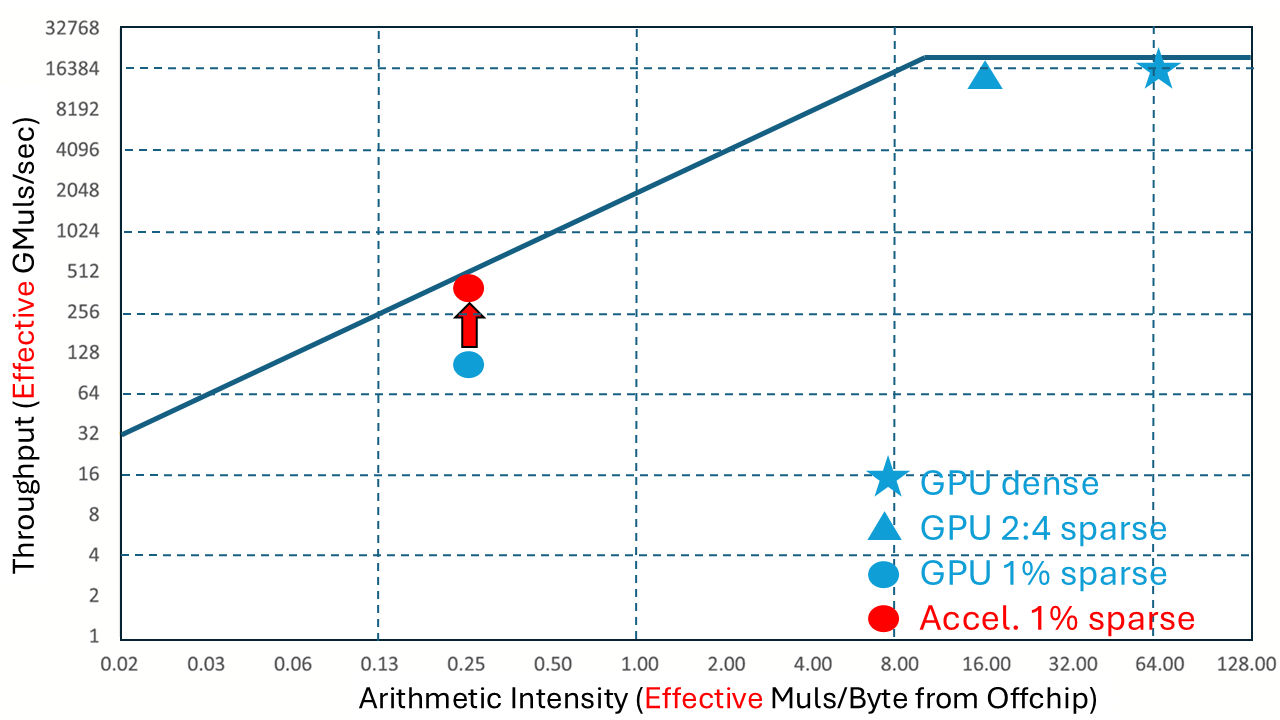}
    \caption{\lb{Structured and unstructured sparsity - Roofline analysis based on data from \cite{pellauer2023symphony}. The red dot corresponds to the Symphony architecture} }
    \label{fig:sparseroof}
\end{figure}

\lb{Analyzing sparse workloads and architectural optimizations on the roofline is non-trivial. Figure \ref{fig:sparseroof}, derived from \cite{pellauer2023symphony}, exemplifies the challenge. All the blue symbols correspond to a parallel tensor contraction kernel executed on the A100 GPU at different levels of sparsity. The dense version of the kernel is very close to the roofline in the compute-bound region, indicating high utilization and architectural efficiency. The version exploiting structured sparsity is still close to the roofline, but slightly more detached and moved to the left (toward the memory-bound region). This is motivated by the significantly reduced number of operations \mv{effectively} executed (only those with non-zero operands in both tensors) and the less aggressive reduction in memory accesses. The latter is caused by the need to load the index of non-zero elements, as well as the fact that non-zero elements in one tensor can be loaded and then not used if they do not correspond to non-zero elements in the other tensor. Furthermore, the reduced burstiness of memory accesses, due to the smaller size of the tensor tiles in compressed form makes it harder to fully exploit the wide transfers at the main memory interface (HBM). Finally, the highly sparse (non-structured) version of the kernel uses indirection, which increases the number of short-burst memory accesses and \mv{decreases the Arithmetic Intensity further. The result is a reduction in} the number of effective GOPS that can be executed per second (because of the extra \mv{memory stalls and} bookkeeping instructions needed). This implies a shift toward the memory-bound region of the roofline and a very significant detachment from it, \mv{due to under-utilization of the peak memory bandwidth}.}

\lb{Architectures with advanced support for sparsity, both hardwired (e.g. \cite{koul2024}) and flexible (e.g. Symphony \cite{pellauer2023symphony}, Occamy \cite{scheffler2025}), corresponding to the red dot in figure \ref{fig:sparseroof}, can significantly reduce the gap from the roofline, thanks to hardware-accelerated indirection. Unfortunately, the kernel remains in the memory-bound region of the roofline because hardware optimization does not change the number of executed products and sum operations with respect to the memory accesses. \mv{The effective Arithmetic Intensity (computed with only the non-zero memory accesses and non-zero operations) remains lower than for the dense workload, signaling less opportunities for data reuse}. Hence, if we look at the roofline alone, it may seem that sparsity exploitation is not a good idea, as it reduces the effective GOPS/sec. However, the very significant reduction of operations may still lead to \mv{absolute energy and latency} 
advantages with respect to ignoring sparsity, using only dense kernels. Roofline analysis alone, albeit useful, is not sufficient to ascertain \mv{the overall} end-to-end \mv{task} speedup and energy efficiency improvements, especially when contemplating the addition of dedicated hardware acceleration for unstructured sparsity. Recent literature results \cite{pellauer2023symphony} indicate that efficiency improvements exceeding one order of magnitude are possible through dedicated hardware if tensors are very sparse (non-zero density below 0.39\%). Current ML models are quite more dense than that. Hence while structured sparsity is immediately exploitable for current models with some tuning effort, non-structured sparsity requires model evolutions toward higher level of sparsity.} 

It is also important to note that quantization and sparsity may be seen as two facets of the same data representation problem: we conjecture that in the future they will be combined in advanced data representation schemes with hardware support aiming at more aggressively reducing the number of bits needed, on average, to store and manipulate weights and activation. On the other hand, the more aggressively quantized are the elements of a tensor, the harder is to get additional advantages by sparsity exploitation (since indexing overhead becomes relatively larger). Optimally balancing sparsity and quantization with efficient hardware, locally and at the level of global memory access, is an interesting and open research challenge.

\subsection{Exploiting near-/in-memory compute} 

\begin{figure}[]
    \centering
\includegraphics[width=\linewidth]{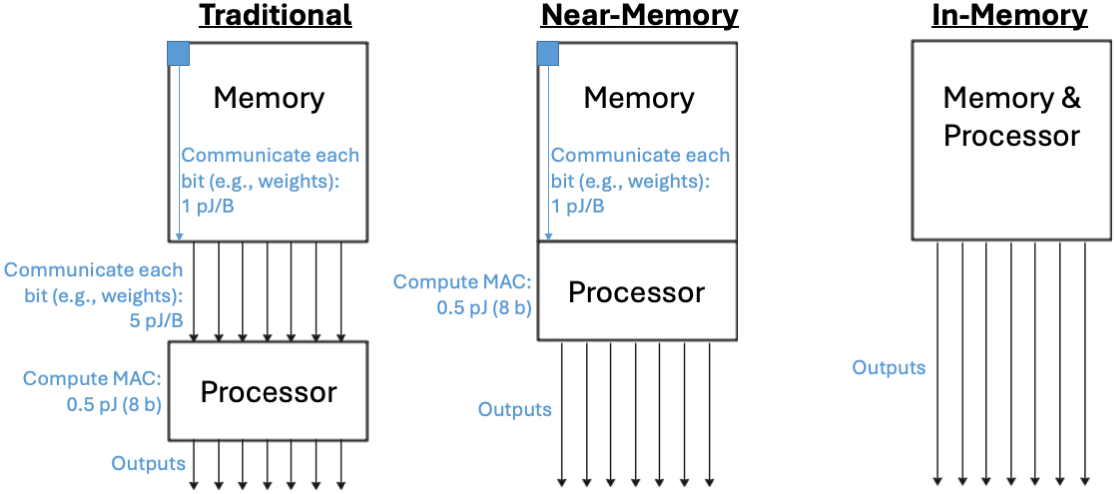}
    \caption{Comparison of traditional, near-, and in-memory architectures }
    \label{fig:tni-mem}
\end{figure}

The need to simultaneously address both the efficiency over a large number of operations and the data-movement costs across large amount of data has motivated near- and in-memory computing architectures. Fig. \ref{fig:tni-mem} shows the progression to such architectures from traditional computing architectures.  

As already considered in detail, the traditional architecture is based on the separation of a large amount of computation from a large amount of memory, with the hardware for each sized to provide the horizontal (compute-bound) and diagonal (memory-bound) rooflines. As we boost these rooflines, the physical size of both hardware structures also scales, leading to substantial data communication costs (energy, latency) during computations. Representative energies for communicating data out of the memory and from the memory to the processor are shown \mv{in Fig. \ref{fig:tni-mem}} for reference. 

The near-memory computing (NMC) architecture is based on finer-grained partitioning and proximate placement of the memory and computation hardware, so that the data-communication energy efficiency and aggregate bandwidth are substantially enhanced. This yields \mv{an} upward shift in the diagonal roofline. Again representative energy numbers are shown for reference. Such an architecture is amenable under two conditions: (1) the computations involve substantial parallelism, so that the computing hardware can be partitioned accordingly; (2) the data required for different parallel computations is well structured and exhibits high locality, such that it can be readily stored in the partitioned memory associated with each parallel computation. As mentioned, both of these conditions are fortunately typical in ML workloads (as well as a range of other signal-processing workloads), making various forms of NMC a natural and commonly employed evolution on traditional integrated architectures. \nv{However, the NMC architecture has also been adopted within memory-centric (e.g., DRAM) technologies, providing integrated computation for data reduction over energy- and bandwidth-limited communication channels, opening new processor-memory architectural options and trade-offs (not in discussed in detail here) \cite{Guseul:asplos24, alsop2024}.}

The in-memory computing (IMC) architecture eliminates the explicit separation of memory and computation hardware, and in doing so eliminates the costs of both communicating data out of the memory and of communicating data from the memory to a processor. In doing so, IMC introduces the potential for substantial efficiency and performance advantages, but through a number of critical circuit and architectural trade-offs that impact computation accuracy and hardware utilization. While NMC is a straight-forward evolution on traditional architectures, IMC presents a number of differentiating considerations, and is the focus of the following section.   

Before analyzing its operation and trade-offs, we note that IMC can be seen as an extreme case of two-dimensional ({2-D}) spatial architectures. Such architectures leverage a 2-D arrangement of PEs, each providing local data storage and computation, in order to exploit 2-D data reuse in matrix operations. In IMC, PEs are reduced to high-density memory bit cells, and the opportunities presented for reducing data communication can be viewed by considering each of the operands involved in a matrix operation. For inputs and outputs (activations) provided to/from the IMC macro, high-density memory bit cells enable greater spatial data reuse, by virtue of a greater number of PEs possible within an allocated area. For Weights stored in the IMC macro, the integration of computation capabilities within bit cells eliminates explicit accessing, at least from a first-level memory, thereby eliminating the roofline diagonal associated with the first-level memory. \nv{As described below, IMC architectures typically require the use of additional level-2 memory for AI model scalability. Thus, viewing it as a high-density PE array (enabling greater parallelism within an allocated area) with enhanced energy efficiency (enabled by analog or custom-digital hardward), IMC has two effects on the roofline model: (1) it boosts the compute-bound roofline; (2) enables increased levels of reuse, moving workload operating points further into the compute-bound regime.} 

\subsection{In-memory compute (IMC) trade-offs and approaches}

A range of circuit approaches have been proposed for enhancing the energy efficiency of computation, and a number of these have been explored within IMCs architectures, as mentioned further below. On the other hand, IMC fundamentally addresses data-movement through a specific approach, namely by performing in-memory data reduction. In doing so, IMC fundamentally instates a dynamic-range trade-off versus energy efficiency and compute density. 

\begin{figure}[]
    \centering
\includegraphics[width=\linewidth]{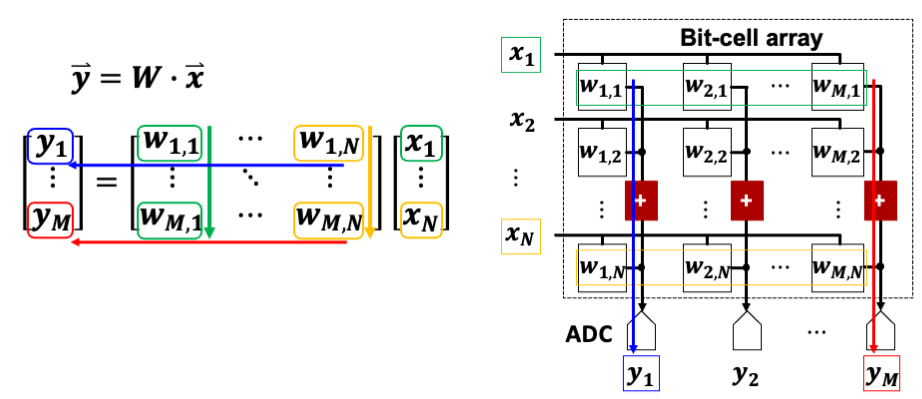}
    \caption{Basics of in-memory computing for MVM operations.}
    \label{fig:in-mem-basics}
\end{figure}

While IMC has been proposed for a range of compute operations, Matrix-Vector Multiplications (MVMs), which are of particular importance in ML workloads, provide the most direct opportunity for such data reduction. This arises because MVMs first involve parallel operations, namely multiplication between input-vector elements and matrix weights, followed by a reduction operation, namely accumulation along the matrix inner dimension. Fig. \ref{fig:in-mem-basics} shows a typical in-memory computing architecture. Input-vector elements are provided in parallel to the memory rows, via the word lines (WLs) or other dedicated inputs, realizing spatial input reuse. The bit cells, either via existing or added circuitry, then perform multiplication with stored data. Finally, accumulation is then performed along each column, on the bit lines or other dedicated outputs, realizing spatial output reuse. 

IMC designs have covered a broad range of approaches to this basic operation. Input data has been provided through different signal-modulation schemes, leveraging pulse-amplitude modulation \cite{zhang_JSSC17} and pulse-width modulation \cite {kang_ICASSP14}. Bit-cell storage has been provided through binary \cite {zhang_JSSC17, kang_ICASSP14} and multi-level \cite{Guo_IEDM17, Fick_CICC17} cells, while multiplication and accumulation has been performed through digital \cite{Fujiwara_ISSCC22, Chih_ISSCC21} and analog \cite{zhang_JSSC17, Guo_IEDM17, valavi201964} circuits. Finally, output readout has been accomplished by direct digital representation and analog-to-digital conversion from the current \cite{Deaville_ESSCIRC21}, voltage \cite{Jia_JSSC20}, and time \cite{Jung_Nature22} domains. In all cases, IMC instates specific fundamental trade-offs, constrained by the storage and compute technologies employed, and mediated by critical design parameters described next.

\subsubsection{IMC Fundamental Trade-offs}

\begin{figure}[]
    \centering
\includegraphics[width=\linewidth]{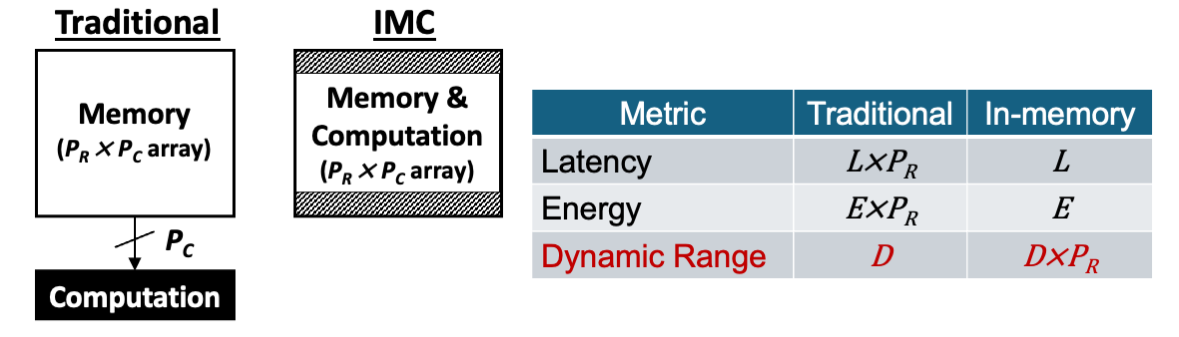}
    \caption{Fundamental dynamic-range trade-off of in-memory computing \cite{Verma_SSCS19}.}
    \label{fig:in-mem-tradeoffs}
\end{figure}

From its basic operation, we see that IMC derives its gains from high-density PEs (bit cells), providing in-memory data reduction. Fig. \ref{fig:in-mem-tradeoffs} analyzes the critical IMC trade-offs determining the achievable gains, compared with traditional/NMC architectures, mediated by the row parallelism $P_R$ (where $P_C$ is the column parallelism). Simplifying the more detailed analysis in \cite{Verma_SSCS19}, we see that a traditional architecture requires energy $E\times P_R$ and latency $L\times P_R$ (from $P_R$ switching cycles), while IMC requires only $E$ and $L$ energy and latency, respectively. However, while a traditional architecture accesses individual bits of stored data having dynamic range $D$, IMC access a compute result via accumulation over $P_R$ bits of data, causing dynamic range increase to at least $D\times P_R$.

Generally, maximizing density and supporting increased dynamic range has energy and performance, as well as, notably, signal-to-noise ratio (SNR) implications, particularly when leveraging analog operation. This has direct impacts on ML model accuracy, as we also previously saw with quantization, and establishes a critical trade-off space for IMC. Quantitatively, the trade-offs depend on the underlying technologies employed for data storage and computation, with implementations thus ranging from low levels of row of parallelism (2-3 rows, where IMC gains are minimal) \cite{Wang_ISSCC19}, moderate levels of row parallelism (10's of rows) \cite{Spetalnick_ISSCC22}, and high levels of row parallelism (1000's of rows) \cite{valavi201964}. We therefore see that the choice of underlying technologies is critical to the gains possible from IMC. 

It is important to note that the discussion above frames the IMC dynamic range trade-off as mediated by the row parallelism $P_R$. However, the dynamic range following in-memory data reduction (accumulation) also depends on the dynamic range of multiplication operands $D_X$ and $D_W$, inputs and weights respective:
\begin{equation}
D_{y} = (2^{D_y}+2^{D_w}-1)\times P_R
\end{equation}

Since multiplication precedes the critical IMC accumulation, its contribution to the dynamic range can be managed outside of the IMC macro, through bit-sliced operation, where multiple input bits are processed in a bit serial manner and multiple weight bits are processed in a bit-parallel manner. This causes energy and throughput scaling as typically expected in multipliers (i.e., linearly with each operand). Though it changes how IMC accumulation impacts SNR, SNR and thus model accuracy ultimately remains limited by the IMC accumulation operation \cite{Jia_JSSC20, Gonugondola_TCAD22}.  

\subsubsection{Technologies and Implementations}

\begin{figure}[]
    \centering
\includegraphics[width=\linewidth]{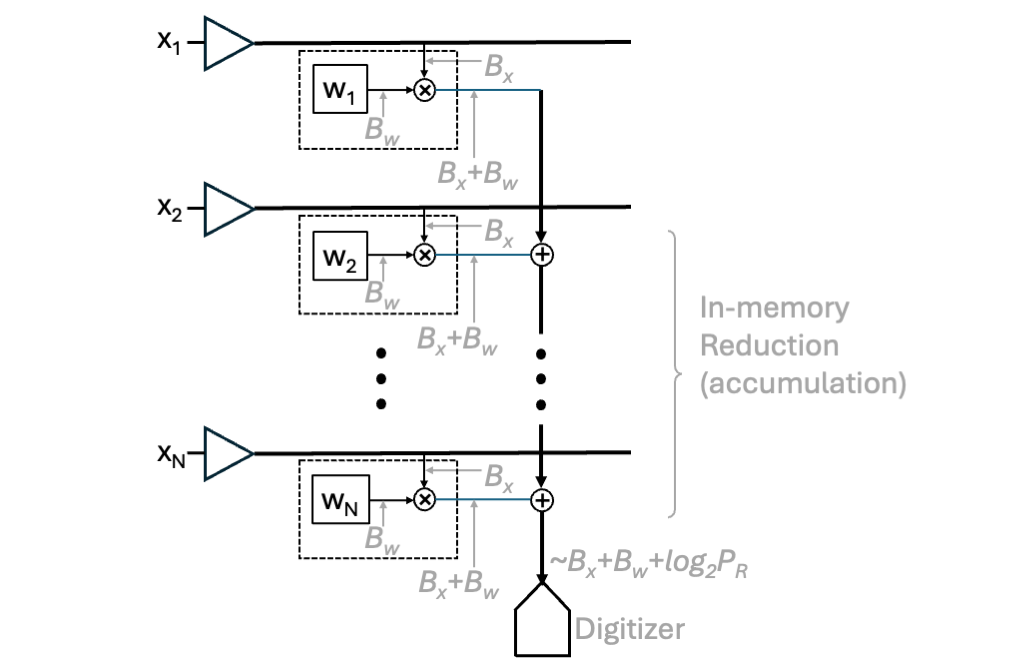}
    \caption{Generalized column computation in in-memory computing.}
    \label{fig:in-mem-col}
\end{figure}

Fig. \ref{fig:in-mem-col} considers a generalized implementation model for an IMC column, comprising: input drivers, for providing row-parallel input data, $X$; bit cells, for storing weight data $W$ and performing multiplication with inputs; mechanism for accumulation; and readout circuitry, for providing digitized outputs. The maximum total dynamic range of the system (in bits) is set by the sum of input bits, $B_X$, and weight bits, $B_W$, involved in each multiplication, plus $log_2$ of the number of rows involved in accumulation $P_R$. The overall dynamic range may be limited by any of IMC circuit components, as discussed below.

\begin{figure}[]
    \centering
\includegraphics[width=\linewidth]{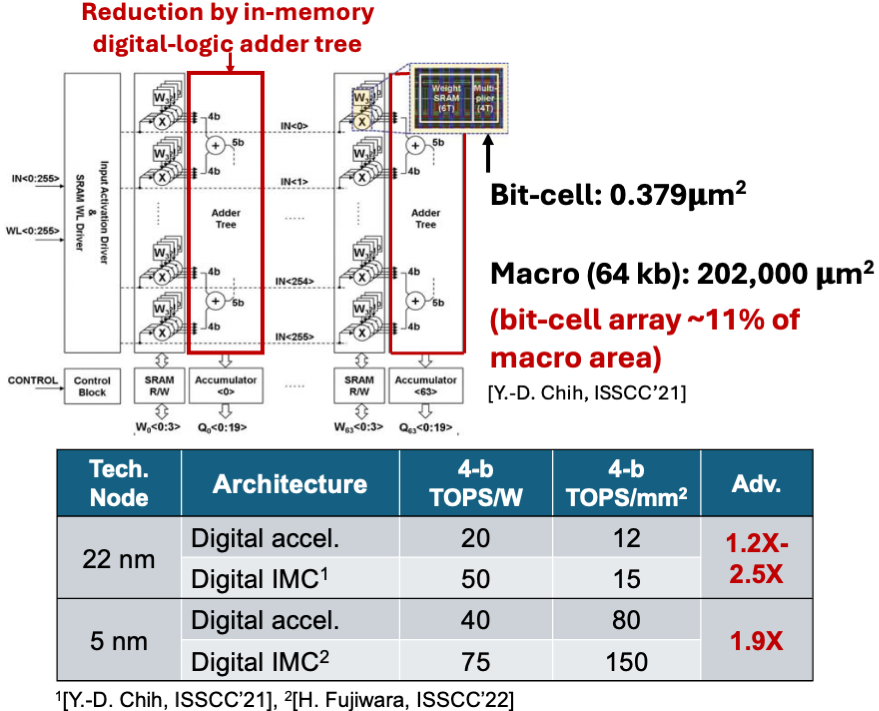}
    \caption{Analysis of digital in-memory computing versus standard digital architectures.}
    \label{fig:in-mem-dig}
\end{figure}

There have been two broad approaches to IMC design: digital IMC (D-IMC), where the critical accumulation operation is performed using an adder tree, based on digital logic, applied across all memory rows \cite{Fujiwara_ISSCC22, Chih_ISSCC21}; and analog IMC (A-IMC), where the critical accumulation is performed using analog current or charge, followed by analog-to-digital conversion \cite{Verma_SSCS19}. Due to the efficiency with which digital circuits can support dynamic-range scaling, D-IMC can readily preserve computation SNR and thus model accuracy. However, by employing the same fundamental technology as digital accelerators, it provides only incremental gains in energy efficiency and compute density relative to traditional digital architectures. As shown in Fig. \ref{fig:in-mem-dig}, D-IMC architectures are typically dominated by their digital circuitry, such that the MVM-level gains achieved relative to traditional digital architectures are in the range $2\times$, arising primarily due to the use of custom-layout adder and weight-storage logic circuits. To reduce the digital circuitry and increase data-storage density, recent D-IMC designs have attempted to time-share the computation logic across multiple sets of bit cells, but where the original compute density and energy efficiency remain largely unchanged due temporal operation \cite{Pascal-ISSCC24}.   

On the other hand, A-IMC has shown the potential for substantially higher energy efficiency than digital architectures \cite{Shanbhag_OJSSCS22}, but has primarily been limited in SNR by various sources of analog noise. Many A-IMC approaches have attempted to scale the operation of typical memory cells, which communicate a stored bit via an output current, in order to minimize the need for additional devices. However, A-IMC implementations based on SRAM \cite{zhang_JSSC17}, flash \cite{Guo_IEDM17}, ReRAM \cite{Spetalnick_ISSCC22}, MRAM \cite{Deaville_VLSI22, Jung_Nature22}, PCM \cite{LeGallo_Nature23} have all shown that the bit-cell current variation and nonlinearity introduce substantial SNR and model-accuracy degradation, preventing scaling from single-bit outputs to the larger dynamic range following accumulation, required for IMC efficiency and throughput advantage.  

Emerging resistive embedded non-volatile memory (eNVM), particularly ReRAM \cite{Chou_ISSCC18} and MRAM \cite{Shum_VLSI17}, which have recently begun to be available in foundry CMOS technologies, have been of particular research interest due to their potential for density scaling in advanced CMOS nodes. However, these technologies tend to provide reduced IMC signal, due to low on-to-off resistance ratios, in the range of 2-10$\times$ (i.e., well below those of MOSFETs used in SRAM technologies, in the range of $10^4\times$), and low absolute resistance. With high IMC row-parallelism, this leads to low SNR and high current for voltage sensing, thus leading to poor readout-circuit power and area efficiency, \nv{often dominating over the memory array itself}. This has prohibitively limited the practical efficiency and compute density achievable by eNVM-based IMC.

\begin{figure}[]
    \centering
\includegraphics[width=\linewidth]{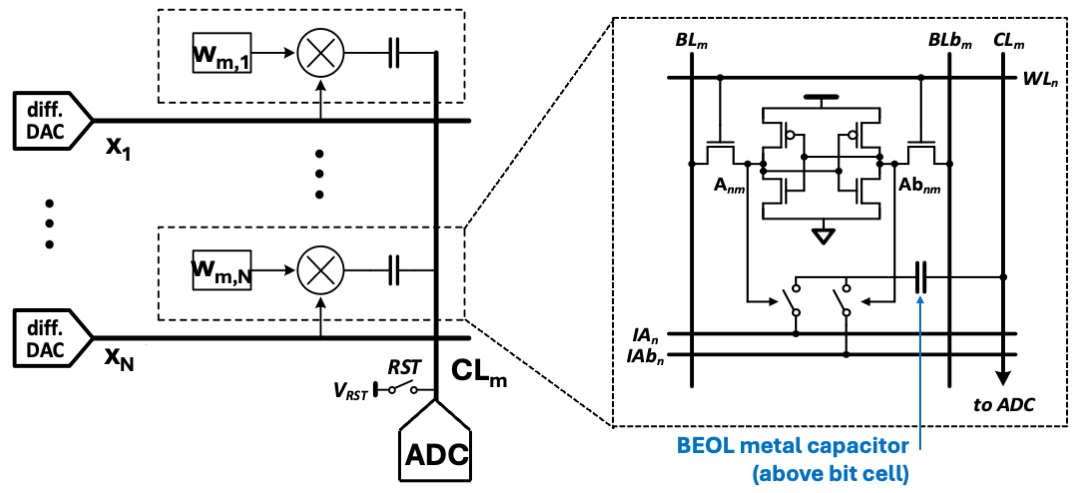}
    \caption{Switched-capacitor in-memory computing for high-SNR computation.}
    \label{fig:in-mem-sc}
\end{figure}

The fundamental SNR tradeoff in IMC, has instead motivated switched-capacitor (SC) operation \cite{valavi201964}, which moves away from current-based signaling of individual bit cells, as prominently used in standard memory operation. As shown in Fig. \ref{fig:in-mem-sc}, SC IMC leverages highly-precise and temperature-stable lithographically-processed backend-of-line (BEOL) metal capacitors, formed above the bit cells, to enable high SNR charge-domain accumulation. Analysis has shown that this can enable row-parallel operation in the tens-of-thousands, with practical demonstration of over 4k-row implementations \cite{valavi201964}. \nv{Further, voltage output signals from SC operation enable direct feeding of energy- and area-efficient ADC architectures (e.g., capacitive SAR), but still impose energy and area overheads ranging from 15-40\% \cite{Jia_JSSC20, Lee_ESSCIRC24}.}

SC IMC has led to complete and scalable full-architecture demonstrations, integrating A-IMC with digital infrastructure (accelerators, CPUs, on-chip networks, level-2 memory) \cite{verma_jssc}, as well as the highest efficiency IMC macros to date (up to 120 TOPS/W for 8-b computation in 28 nm CMOS) \cite{Lee_ESSCIRC24}.

\subsubsection{Architectural Implications}

At the architectural level, a critical challenge introduced by IMC is that it intrinsically couples storage and compute resources, yet maximizing utilization and scalability requires optimizing each of these separately. \nv{This prevents a fully-weight-static workload mapping. To illustrate Table \ref{table:imc-util-store}, considers fully-weight-static mappings optimizing storage utilization and compute utilization, respectively, while analyzing the isolated effect of utilization losses that result from the conflict between these (ignoring all other sources of utilization loss, i.e., assuming perfect input activation feeding). Columns 3-5, consider workload mappings that optimize for storage, by storing each of the model weight bits statically in only one IMC bit cell. Since weights are generally involved in highly differing number of operations depending on the model architecture, this can lead to extremely low compute utilization, due to temporal utilization loss in IMC bit cells storing weights involved in fewer operations. As shown, while transformers (e.g., BERT), comprised primarily of dense layers, maintain uniform number of operations for all weights in the model, and thus preserve high compute utilization, convolutional backbones, including ones used in generative vision models (e.g., UNet-2D), exhibit widely varying operations, thus substantially limiting compute utilization across workloads. Conversely, columns 6-8 consider workload mappings that optimize for compute utilization, by replicating weight bits in IMC bit cells in proportion to the number of operations. This requires to the excessive number of IMC bit cells shown, effectively leading to low spatial storage utilization across workloads, thus substantially limiting architectural scalability.} 

\begin{table}[]
    \centering
    \caption{\nv{Analysis of IMC, optimizing for storage vs. compute utilization in a fully-weight-static mapping, considering utilization losses just due to the conflict between these.}}
    \includegraphics[width=\linewidth]{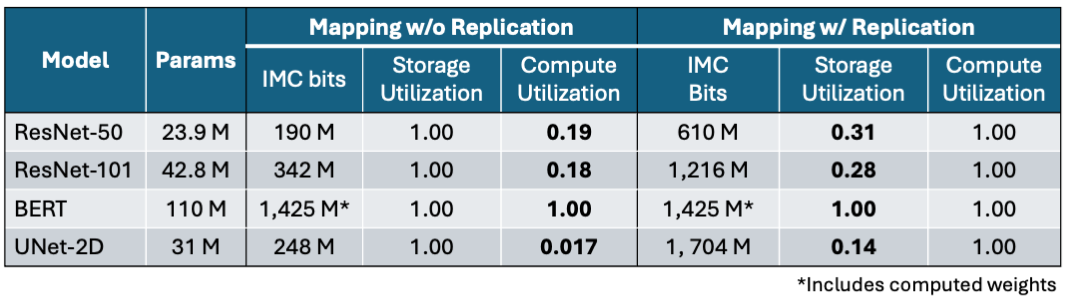}
    \label{table:imc-util-store}
\end{table}

Such analysis illustrates the need to ultimately break the compute-storage coupling through the use of level-2 (L2) memory, so that workload computations can be temporally mapped, to address architectural scalability while ensuring high utilization. This has important implications at the level of IMC circuit and technological design, as well as at the level of architectural design. In terms of circuit and technological design, for instance, IMC write energy and bandwidth become critical for efficient temporal mapping of workloads, limiting the applicability of eNVM technologies, due to their write energies and endurances. In terms of architectural design, the need for efficient spatial (parallel) and temporal mapping of workloads invokes the many considerations discussed earlier in this paper, but now optimized to the distinct circuit and microarchitectural attributes of IMC, in terms of area, compute throughput/energy, and weight-loading bandwidth/energy. Architectural research in IMC has begun to consider such factors. Doing so has exposed the overheads of temporal execution and different forms of parallel execution (data, model, pipeline), which reduce the achievable compute rooflines, and also motivated specialized architectures to minimize such overheads \cite{verma_jssc}.

\section{\nv{Roofline models for trade-space exploration}}

\nv{The roofline model represents salient aspects of an architecture, illustrated by the various ways the architectural techniques discussed in Section \ref{sec:techn} either raise the performance or energy rooflines, or enable workloads to more closely approach the performance or energy rooflines. This implies that, in addition to characterizing a given architecture, the roofline model can serve as a useful tool for architectural design-space exploration and analysis. This section examines how the roofline model can drive architectural decisions, by considering how different architectural trade-offs impact the roofline model. Ultimately, the workload-level performance and efficiency across a relevant set of workloads, would be determined by combining the roofline model with the workload mapping, which impacts the arithmetic intensity and utilization.}

\nv{As it is not straightforward for a designer to find the best trade-off between these conflicting design considerations, several performance modeling and design-space exploration frameworks have been appearing in the state-of-the-art~\cite{Interstellar, MAESTRO, Timeloop, ZigZag, CoSA, MindMap, GAMMA, Stream}. These frameworks allow to model many hardware architecture variants and can estimate the execution cost of running specific workloads on them, while optimizing the efficiency of the spatial and temporal mapping through loop optimizations such as unrolling, tilling and reordering. Nonetheless, roofline models provide a valuable tool for building intuition and rationale underpinning the results of such analysis frameworks.}

\subsection{Compute versus memory area allocation} 

\begin{figure}[]
    \centering
\includegraphics[width=\linewidth]{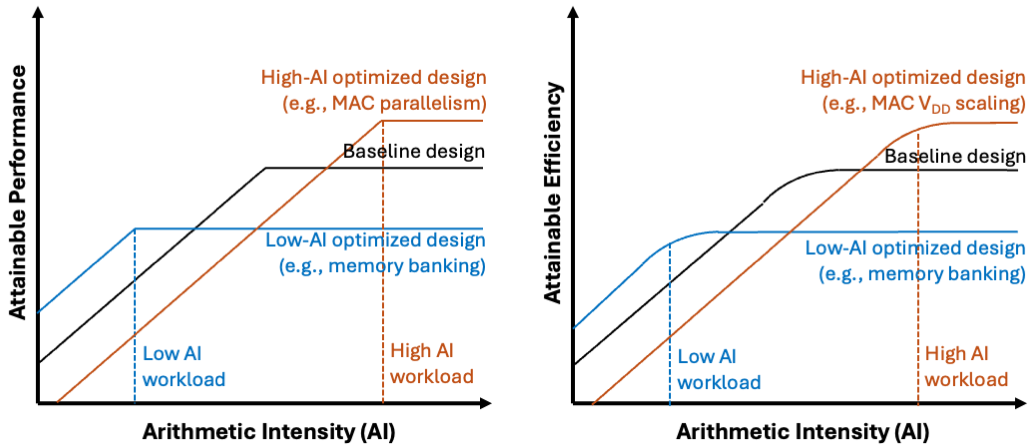}
    \caption{Roofline optimizations to different workload arithmetic intensities.}
    \label{fig:roofline-opt}
\end{figure}

The methods previously suggested for raising the compute-bound and memory-bound rooflines directly impact area and energy, i.e., MAC parallelism and voltage scaling in the case of compute-bound performance and efficiency, and memory banking in the case of memory-bound performance and efficiency. This suggests an approach to architectural optimization under a joint area and energy 
constraint. As shown in Fig. \ref{fig:roofline-opt}, the level of compute parallelism and memory parallelism (banking) can be optimally set to the arithmetic intensity presented by a target workload. The critical challenge, in practice is that architectures must typically be designed to support a range of workloads with differing arithmetic intensities, and where programmability overheads and associated utilization losses will further reduce actual efficiency and performance.  

Of recent relevance, the explosion in ML model size has made the total memory capacity required a critical challenge even at the level of off-chip DRAM, thus extending critical memory-bandwidth and compute-parallelism trade-offs beyond a single chip. For instance, the capacity limitations of contemporary DRAM high-bandwidth memory (HBM) necessitate executing Large-Language-Models (LLMs) across multiple chips, each with attached HBM. This can require interconnection between multiple (16-32) chips, making inter-chip networking an additional key technology, together with compute and memory technologies. In fact, to simultaneously meet memory bandwidth and capacity requirements, architectures have been proposed based only on high-bandwidth on-chip SRAM, integrating several hundreds of MB's on chip and requiring over 500 chips to be networked even for LLMs of modest size \cite{Abts_HotChips22}.

\subsection{Parallelism 
versus utilization}

\nv{While the previously discussed approach of employing large processing arrays raises the rooflines by maximizing parallelism and minimizing the energy per operation, the challenges inherent with spatial, temporal, and core utilization presented in Section \ref{sec:tech-parallel} can often prevent operation near the rooflines. To analyze workload-level performance the roofline model must be considered together with the workload operating point, which itself is determined by the workload mapping. As we will see, architectural decisions favorably impacting the rooflines can adversely impact the workload operating point.}

\nv{For instance, larger spatial arrays make it much harder to maintain full spatial utilization across a wide range of workloads. For example, if we map a workload with only 128 output channels on the analog IMC core of Figure \ref{fig:davinci} (bottom), only 32,768 of the 65,536 MACs will be activated during a typical compute cycle, resulting in a spatial utilization of just $50\%$.} 

\nv{Additionally, large monolithic processing arrays require a massive amount of data per clock cycle, limited by memory bandwidth. The severity of such an effect depends on the workload mapping, where leveraging greater levels of data reuse enhances arithmetic intensity, pushing back to the compute-bound regime. Larger arrays may enable increased data reuse by virtue of storing more data involved in the operations, but the maximum level of data reuse is ultimately limited by the workload itself.}   
For example, if we have to reload the weights of the IMC array of \ref{fig:davinci} (bottom) every 1024 cycles, and this weight loading takes 256 cycles (1 row per cycle) during which no compute can take place, 1/5th of the clock cycles is lost to weight transfers resulting in peak temporal utilization of just $80\%$. 
The total utilization per processor core \mv{(the product of the spatial and the temporal utilization)}, determines the distance between the roofline peak performance and the obtained performance within that core. Utilization hence puts a limit on how close the processor can approach the roofline performance (Figure~\ref{fig:RLimpr} (right)).

These tradeoffs become particularly critical with IMC architectures, due to their parallelism tradeoffs and relative efficiency of compute versus weight transfers. In terms of parallelism, as previously discussed, IMC derives its efficiency advantages from high-levels of row parallelism and compute-density advantages from integration of many dense bit cells within each core, making spatial (bit cell) utilization loss a key challenge. In terms of weight transfers, while row-parallel data reduction improves compute efficiency and throughput in IMC, weight loading remains limited to bit-by-bit writing, making temporal utilization losses of increased severity. Initial demonstrations of IMC architectures have begun to address these challenges, by noting that different forms of parallel execution (data/model/pipeline-parallelism) yield different trade-off points, thus integrating low-overhead supports to optimize across different forms of parallel execution \cite{verma_jssc}.

 The fundamental trade-off stems from the fact that larger monolithic PE arrays (e.g. large IMC arrays or widely-parallel digital tensor cores operating at reduced precision) are good from a roofline point of view, but challenging from an utilization point of view. 
Smaller tensor cores, on the other hand, are easy to keep well utilized both spatially and temporally. Yet, they can only exploit limited data reuse with low arithmetic intensity, resulting in a lower peak array performance. Such a system would have a lower roofline, but which can be approached quite closely across more workloads.

\nv{Similarly, architectural supports for sparsity inevitably cause overheads in the computational and memory-accessing control flow, yielding reduced rooflines. However, when sparse operations are analyzed as no-operations, such architectures yield substantially higher utilization compared to architectures with no sparsity support, where the no-operations directly result in utilization loss. In this way, roofline models expose the benefits of low overhead approaches to sparsity support within an architecture (as discussed in Section \ref{sec:sparsity} \mv{and Figure 15}), by minimally reducing the rooflines while enabling high workload utilization (operating points near the rooflines).}

\subsection{Flexibility (programmability) versus specialization}
All the previous sections have described approaches that involve some degree of tailoring hardware to the ML application domain. Hence, the game of increasing efficiency is mostly a game of specialization. The multi-order-of-magnitude efficiency boost associated with domain specialization has been known for a long time, well before the ML hardware explosion \cite{Gotz06,Huang22}. However, there are two unique differentiating factors associated with AI that favor specialization. First, the application markets are huge, leading to large potential production volumes and margins, which can justify the non-recurring engineering cost of a specialized architecture. Second, most  ML models are dominated, from a computational standpoint, by relatively few kernels, namely, linear and few non-linear tensor operators \cite{Ghodrati24}.  Furthermore, for all deep neural models the arithmetic precision needed for these operators can be much reduced, as seen in section \ref{sec:quant}.  The combination of these factors has led to a "specialization gold rush" with a proliferation of ultra-specialized accelerators.

It is important to note however, that even in this unique context, hyper-specialization is a dangerous slope, with the pitfall that aggressively specialized NPU architectures may achieve the lowest energy-per-operation at peak utilization, but are subject to under-utilization when workloads change and evolve \cite{Ghodrati24}. Hence, a poorly utilized, ultra-specialized NPU can become less efficient at the workload level than a highly utilized, but less specialized, architecture (e.g. a GP-GPU). If we visualize workloads as points on the roofline plot we ideally would like our architecture to have as many points as close as possible to the roofline. 

A concrete example of such workload evolution is the recent transition of leading-edge models from convolutional neural networks to attention-based networks (transformers). From the computational standpoint the key operator in CNNs, namely the N-channel 2-D convolution, is a high-locality kernel exemplified in Figure \ref{fig:loops}. Convolution can be transformed in general Matrix multiplication through a Toeplitz transformation, with a non-negligible memory inflation \cite{Cho17}. Hence, several acceleration engines have been designed for executing convolutions natively, with no storage overhead. However, the now widespread transformers are dominated by matrix multiplication in the attention block, thus hardware that is specifically designed for convolutions faces critical limitations. 

\mv{Achieving a high utilization across workloads, including unknown future ML workloads, will hence require flexible architectures, capable of adjusting at compile-time or run-time to the workload characteristics. Although it is difficult to quantify flexibility, without knowing the complete kernel space of interest, recent work defined 4 possible axis of flexibility, denoted as TOPS\cite{kao2022formalism}: Tile size (= sizes of the temporal loops); Order (= relative ordering of the temporal loops); Parallelization (= which dimensions to unroll in the spatial loops); and the Shape (= sizes of the spatial loops). These metrics allow to compare the flexibility of existing and envisioned accelerators in terms of their GeMM support.}  

\mv{Workloads can, however, evolve in other directions than purely in terms of their GeMM operations with new alternative operators popping us. This leads us to } another significant risk factor in specialization linked to the well-known "Amdhal effect": if a fraction of a workload $f<1$  cannot be accelerated, the end-to-end speedup that can be achieved by the best ideal accelerated architecture is upper-bounded by $1/f$. Hence, to achieve high end-to-end speedup, a NPU has to be flexible enough to accelerate a dominant fraction of the computational kernels in a neural network, to ensure that $f<<1$. If these kernels are not homogeneous, multiple "sub-accelerators" may be needed within the NPU. Dimensioning to ensure high utilization for all sub-accelerators, as required to achieve high efficiency, is a difficult balancing act, since silicon area is always limited. Several architectures are being explored to address the accelerator utilization challenge \cite{Ghodrati24}, and this is an active area of research: the sweet spot between efficiency and flexibility depends not only on workload, but also on technology parameters, such as leakage per unit-area, which grows with technology scaling, penalizing under-utilization not only from the area, but also from the power viewpoint. 

\nv{In terms of the roofline model, flexible architectures incur the typically substantial overhead of control flow optimized for software extensibility. This has the impact of reducing both the compute and memory-accessing rooflines. The benefits of such architectures, on the other hand, become apparent from the utilization achieved across a diverse set of workloads (i.e., workload operating points near the roofline). Since the architectural overheads of software-optimized control flow are typically large, judicious selection of the target workload set is often necessary, making ML accelerator design an evolving and application-driven exercise.}


\


\section{Conclusion and open challenges: How to keep pushing ML accelerator performance?}

The past ten years has seen dramatic enhancements in ML acceleration, amounting to roughly 1000$\times$ increase in throughput and energy efficiency. These have come from two synergistic sources: (1) hardware improvements, at the technological, circuit, and architectural levels; (2) algorithmic improvements, in the form of number formats and model architectures, which have in turn been leveraged to their full extent by the integration of hardware supports. 

Today, we see that both of these are hitting limits, bound in one case by the fundamental underlying technologies and in the other case by ultimate task-level accuracies. This has placed ML accelerator design squarely in a regime of balancing across critical trade-offs, spanning peak efficiency, peak throughput, achievable utilization, and workload flexibility. This has made it essential to understand the inter-related trade-offs, in order to effectively design ML accelerators for the diverse range of system deployments being envisioned. \mv{A key tool for reasoning about these trade-offs has been the roofline model, which offers insight into how compute performance and memory bandwidth constraints shape accelerator efficiency. As ML workloads diversify, accelerator architectures must be designed to not only push peak performance (raising the rooflines) but also ensure high utilization across a range of workloads (operating near the rooflines).}

Looking beyond today, the rich range of research makes a number of emerging technologies, across computation, memory, and networking, as well as emerging algorithmic and model architecture innovations, likely to continue to fuel enhancements in ML acceleration. Indeed, maximally leveraging these within future ML accelerators, at the rapid pace with which they are emerging, also makes it essential to understand the trade-offs covered in this paper.    

While a complete survey of emerging technologies is beyond the scope of this paper, a few examples can help to illustrate how such technologies may influence the trade-offs discussed. For instance, chiplets, and their associated technologies (hybrid integration, die-to-die interconnects), will provide promising pathways for scaling compute parallelism and memory bandwidth, \mv{and as such raise the rooflines}. This can be further extended to highly-distributed systems, across many compute and memory nodes, where next-generation interconnect technologies, such as optical communication. Also leveraging emerging memories can provide greater density and enables scale-up of memories to address capacity requirements with ever-increasing model sizes. \mv{Yet again, the challenge is not only to make the roofline as high as possible, but also to operate close to it. Fortunately,} such scale out could also raise data reuse opportunities and introduce arithmetic intensity relaxations between nodes and clusters, which may make new and existing memory technologies relevant, in repurposed forms and with optimized interfaces.

In addition to interconnect and memory technologies, new computing technologies will also play a key role. Analog computation, particularly in specific forms of in-memory computing, is already showing direct relevance for next-generation systems. Its potential is currently being explored on a number of other fronts, including alongside new device and materials innovations.

\mv{As mentioned, however,} in complement to hardware innovations, maximizing utilization of architectures, will remain an important challenge and point of system leverage. This will be achievable by more tightly integrating accelerator performance models with compiler technology. Rapid development cycles in both hardware and software suffer from the current need to micro-code kernels in order to achieve satisfactory utilization. In this context, MLIR \mv{(Multi-Level Intermediate Representation\cite{lattner2021mlir})}, \nv{can form an effective abstraction framework for the development and integration of tools towards rapid processor emulation and code tuning}

In the end, designing hardware for machine learning will always remain a balancing game: Maximizing total system efficiency and throughput, while maintaining sufficient flexibility to be prepared for the continuous stream of innovations in ML algorithms. 
The danger is that new powerful algorithms are of such different nature, that they do not match well on today's processor architectures, preventing their breakthrough. To avoid this hardware lottery \cite{hooker2021hardware}, we need to find a way to not only streamline the design and programming of novel ML accelerator platforms, but also their fabrication. The recent trend towards using chiplet technology not only for achieving scale-out with good yield, but also toward rapid customization seems promising. Future will tell whether this will allows us to again enable the next wave of ML algorithms.


\bibliographystyle{IEEEtran}

\begin{IEEEbiography}
    [{\includegraphics[width=1in,height=1.25in,clip,keepaspectratio]{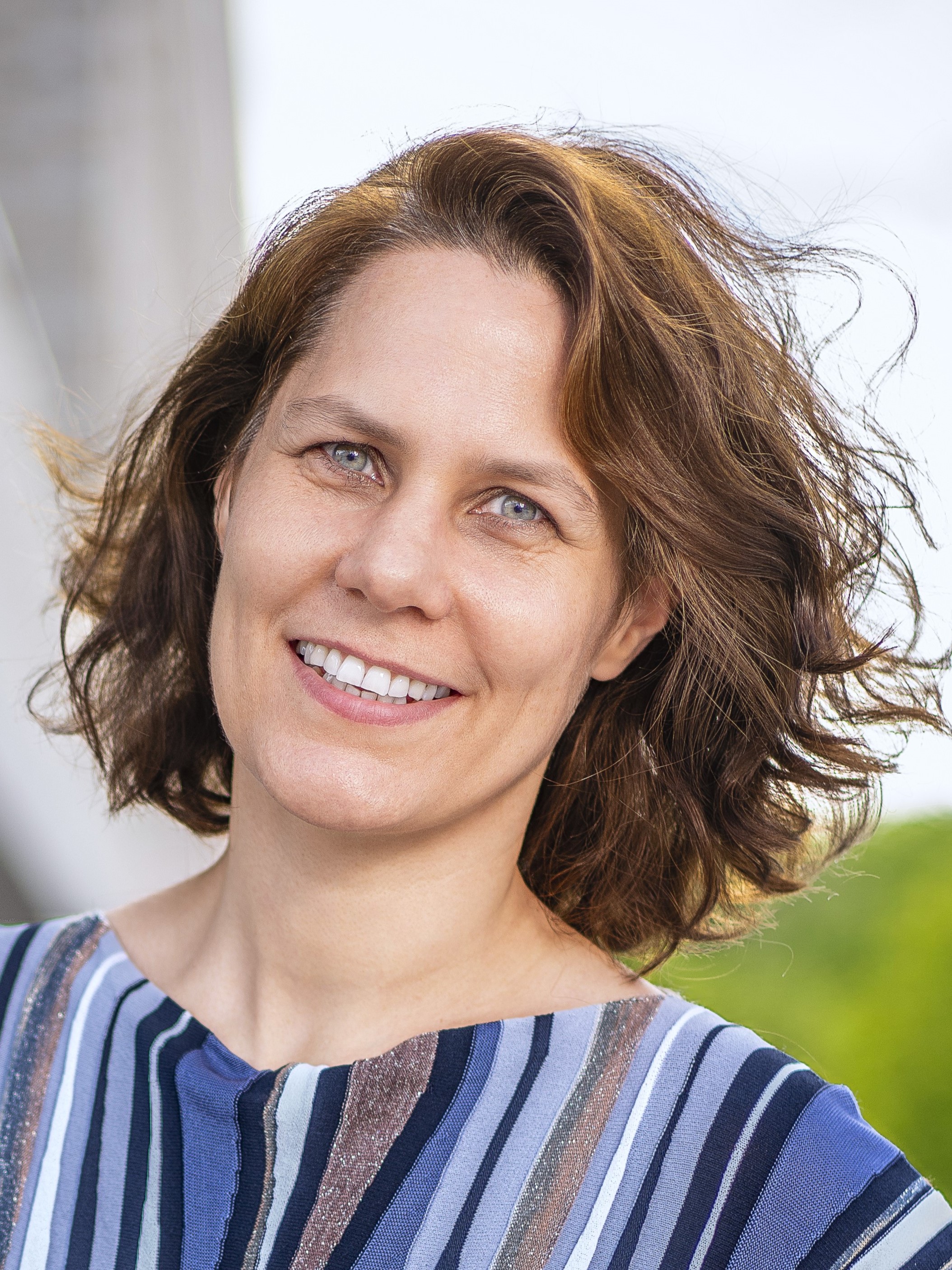}}]{Marian Verhelst}
Marian Verhelst is a professor at the MICAS labs of KU Leuven and a research director at imec. Her research focuses on embedded machine learning, hardware accelerators, and low-power edge processing. She received a PhD from KU Leuven in 2008, and worked as a research scientist at Intel Labs from 2008 till 2011. Marian is a scientific advisor to multiple startups, member of the board of ECSA, member of the Royal Academy of Belgium for Science and Arts, and active in the TPC’s of ISSCC, ISCA and ESSERC. She served in the board of directors of tinyML, as a member of the Young Academy of Belgium, as an associate editor for TVLSI, TCAS-II and JSSC, and as a member of the STEM advisory committee to the Flemish Government. She is a science communication enthusiast as an IEEE SSCS Distinguished Lecturer, as a regular member of the Nerdland science podcast (in Dutch), and as the founding mother of KU Leuven’s InnovationLab high school program. Marian received the laureate prize of the Royal Academy of Belgium in 2016, the 2021 Intel Outstanding Researcher Award, and the André Mischke YAE Prize for Science and Policy in 2021.
\end{IEEEbiography}

\begin{IEEEbiography}
    [{\includegraphics[width=1in,height=1.25in,clip,keepaspectratio]{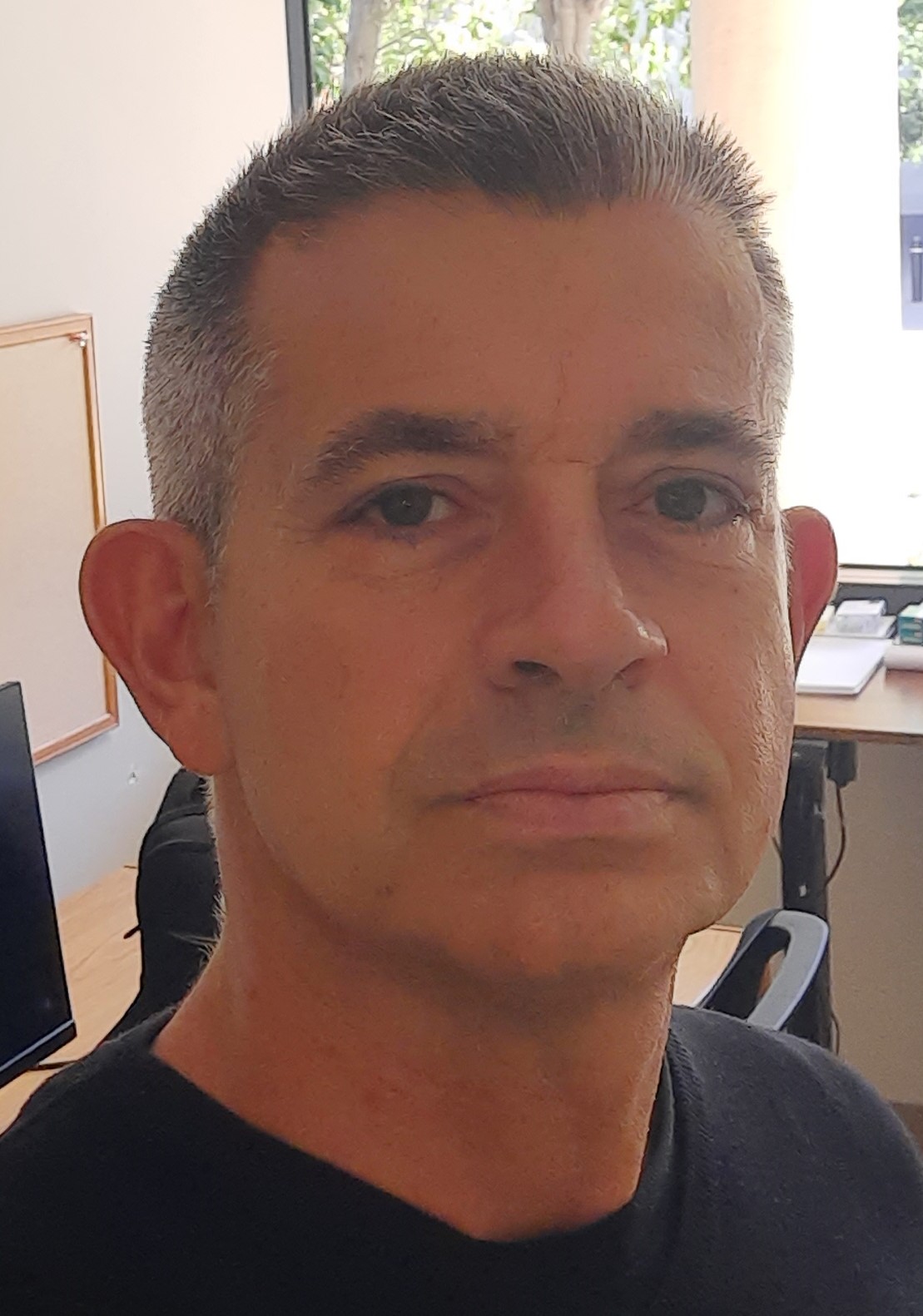}}]{Luca Benini}
Luca Benini holds the chair of digital Circuits and systems at ETHZ and is Full Professor at the Università di Bologna. He received a PhD from Stanford University. His research interests are in energy-efficient parallel computing systems, smart sensing micro-systems and machine learning hardware. He is a Fellow of the IEEE, of the ACM, a member of the Academia Europaea and of the Italian Academy of Engineering and Technology. He is the recipient of the 2016 IEEE CAS Mac Van Valkenburg award, the 2020 EDAA achievement Award, the 2020 ACM/IEEE A. Richard Newton Award, the 2023 IEEE CS E.J. McCluskey Award, and the 2024 IEEE CS Open Source Hardware contribution Award.
\end{IEEEbiography}

\begin{IEEEbiography}
    [{\includegraphics[width=1in,height=1.25in,clip,keepaspectratio]{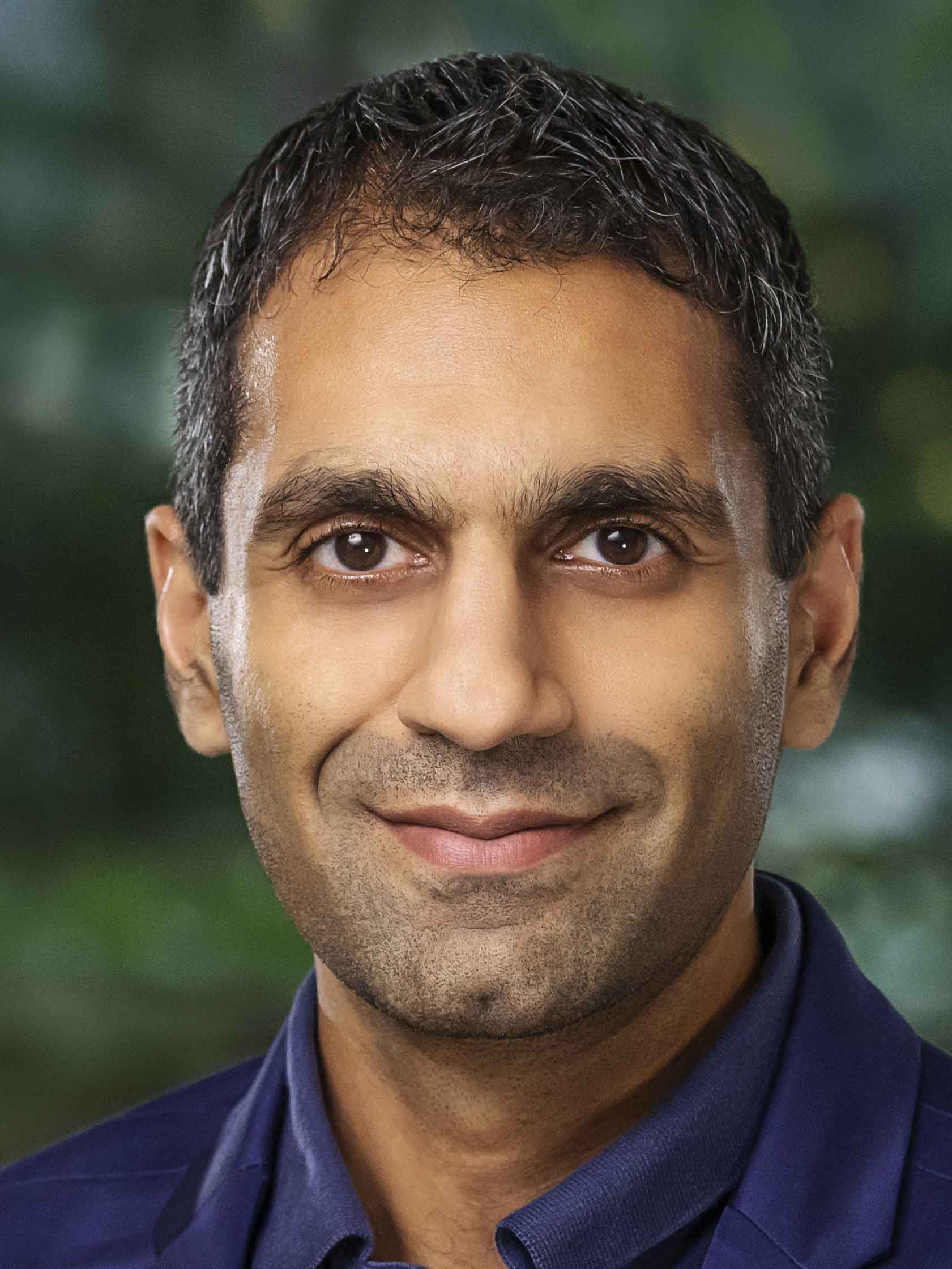}}]{Naveen Verma}
Naveen Verma received the B.A.Sc. degree in Electrical and Computer Engineering from the UBC, Vancouver, Canada in 2003, and the M.S. and Ph.D. degrees in Electrical Engineering from MIT in 2005 and 2009 respectively. Since July 2009 he has been at Princeton University, where he is currently the Ralph H. and Freda I. Augustine Professor of Electrical and Computer Engineering. His research focuses on advanced sensing and computing systems, exploring how systems for learning, inference, and action planning can be enhanced by algorithms that exploit new sensing and computing technologies. This includes research on large-area flexible sensors, energy-efficient computing architectures and circuits, and machine-learning and statistical-signal-processing algorithms. Prof. Verma has been involved in a number of technology transfer activities including founding start-up companies. Most recently, he co-founded EnCharge AI, together with industry leaders in AI computing systems, to commercialize foundational technology developed in his lab. Prof. Verma has served as a Distinguished Lecturer of the IEEE Solid-State Circuits Society, and on a number of conference program committees and advisory groups. Prof. Verma is the recipient of numerous teaching and research awards, including several best-paper awards, with his students.
\end{IEEEbiography}

\vfill

\end{document}